\documentclass{pasa}

\usepackage{amsmath}
\usepackage{amsfonts}
\usepackage{amssymb}
\usepackage{graphicx}
\usepackage{mathrsfs}
\usepackage{latexsym}
\usepackage{graphics}
\usepackage{mathtools}
\usepackage[english]{babel}
\usepackage{bm}
\usepackage{color}
\usepackage{xspace}
\usepackage{listings}

\usepackage[authoryear]{natbib}
\bibpunct{(}{)}{;}{a}{}{,}
\newcommand{\mathbfss}[1]{\mathbf{#1}}
\def\be{\begin{equation}}
\def\ee{\end{equation}}
\def\ben{\begin{equation*}}
\def\een{\end{equation*}}
\def\na{\begin{align}}
\def\ea{\end{align}}
\def\x{\mathbfss{x}}

\def\xx{\bm{\xi}}
\def\k{\mathbfss{k}}
\def\sxyi{\sigma_{\!xy,i}}
\def\sxi{\sigma_{\!x,i}}
\def\syi{\sigma_{\!y,i}}
\def\ci{c_{i}}
\def\t{^\top}
\def\a{\bm{\alpha}}

\def\C{{\mathbfss C}}
\def\n{{\mathbfss{n}}}
\def\RD{{\mathbb{R}^D}}
\def\H{\mathcal{H}}

\def\Z{\mathbb{Z}}
\def\O{\mathcal{O}}
\def\L{\mathcal{L}}
\newcommand\hf{\textsc{hyper-fit}\xspace}
\newcommand\ld{\textsc{laplacesdemon}\xspace}
\newcommand\R{{\textsc R}\xspace}
\newcommand\Shiny{{\textsc Shiny}\xspace}
\newcommand\JAGS{{\textsc JAGS}\xspace}
\newcommand\BUGS{{\textsc BUGS}\xspace}
\newcommand\Stan{{\textsc Stan}\xspace}

\date{\today}

\title[Hyper-Fit]{Hyper-Fit: Fitting Linear Models to Multidimensional Data with Multivariate Gaussian Uncertainties}
\author[A.S.G. Robotham \& D. Obreschkow]{
A.S.G.~Robotham$^1$ \& D. Obreschkow$^1$\\
\affil{$^1$ICRAR, M468, University of Western Australia, Crawley, WA 6009, Australia}
}

\begin{document}




\begin{abstract}

Astronomical data is often uncertain with errors that are heteroscedastic (different for each data point) and covariant between different dimensions. Assuming that a set of $D$-dimensional data points can be described by a $(D-1)$-dimensional plane with intrinsic scatter, we derive the general likelihood function to be maximised to recover the {\it best fitting} model. Alongside the mathematical description, we also release the \hf package for the \R statistical language (github.com/asgr/hyper.fit) and a user-friendly web interface for online fitting (hyperfit.icrar.org). The \hf package offers access to a large number of fitting routines, includes visualisation tools, and is fully documented in an extensive user manual. Most of the \hf functionality is accessible via the web interface. In this paper we include applications to toy examples and to real astronomical data from the literature: the mass-size, Tully-Fisher, Fundamental Plane, and mass-spin-morphology relations. In most cases the \hf solutions are in good agreement with published values, but uncover more information regarding the fitted model.

\end{abstract}


\maketitle

\section{Introduction}\label{section_introduction}

\begin{figure}
	\centering\vspace{7mm}
	\includegraphics[width=\columnwidth]{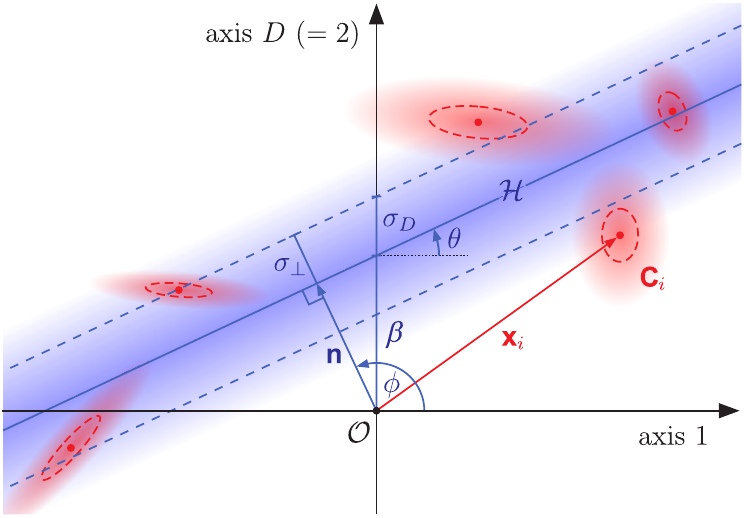}
	\caption{Schematic representation of the linear model (blue) fitted to the data points (red). Both the model and the data are assumed to have Gaussian distributions, representing intrinsic model scatter and statistical measurement uncertainties, respectively. In both cases, $1\sigma$-contours are shown as dashed lines.}
	\label{fig_schema}
\end{figure}

There is a common conception that the method used for fitting multidimensional data is a matter of subjective choice. However, if the problem is mathematically well-defined, the optimal fitting solution is, in many cases, unique. This paper presents the optimal maximum likelihood fitting solution (with the caveats that a maximum likelihood fitting approach entails) for the frequently encountered case schematised in Figure~\ref{fig_schema}, where $D$-dimensional data (i) have multivariate Gaussian uncertainties (these may be different for each data point), and (ii) randomly sample a ($D-1$)-dimensional plane (a line if $D=2$, a plane if $D=3$, a hyperplane if $D\geq4$) with intrinsic Gaussian scatter. In particular, the data have no defined predictor/response variable. Hence standard regression analysis does not apply, and in general it would yield different geometric solutions depending on axis-ordering \citep[though see][which presents the necessary prior modifications to make traditional intrinsic-scatter-free regression analysis rotationally invariant]{vont15}.

Within the assumptions above, the ($D-1$)-dimensional plane that best explains the observed data is unique and can be fit using a traditional likelihood method. In this paper we present the general $D$-dimensional form of the likelihood function and release a package for the \R statistical programming language (\hf) that optimally fits data using this likelihood approach. We also release a user-friendly web interface to run \hf online and apply our algorithm to common relations in galaxy population studies.

Other authors have investigated the problem of regression and correlation analysis \citep{kell07,hogg10,kell11,vont15} with the aim of improving on the ad hoc use of standard linear or multi-variate regression, reduced $\chi^2$ analysis, computationally expensive (and usually unnecessary) bootstrap resampling, and the `bisector method' that is sometimes wrongly assumed to be invariant with respect to an arbitrary rotation of the data. Of recent work that is familiar in the astronomical community, \citet{kell07} presents a comprehensive approach for linear regression analysis with support for non-detections, and made available {\textsc IDL} code that can analyse multi-dimensional data with a single predictor variable. More recently \citet{hogg10} describes the approach to attacking such {\it joint regression} problems in two dimensions. \citet{tayl15} describes an approach to extend the work of \citet{hogg10} to multiple Gaussian mixture models to model the {\it red} and {\it blue} populations in the Galaxy And Mass Assembly survey \citep{lisk15}. Here we extend \citet{hogg10} to arbitrary dimensions and include a suite of tools to aid scientists with such analysis.

In Section \ref{section_deriv} we describe the likelihood method for an arbitrary number of dimensions and for different projection systems (to allow for maximum flexibility in its application). In Section \ref{section_hyper.fit} we describe and publicly release the \hf package for \R, a user friendly and fully documented package that can be applied to such problems. We also introduce a simple web interface for the package. In Section \ref{section_mockexamples} we show simple applications of \hf to toy examples, including the dataset presented in \citet{hogg10}. In Section \ref{section_astroexamples} we show more advanced examples taken directly from published astronomy papers, where we can compare our fitting procedure against the relationships presented in the original work. In the appendix we add a discussion regrading the subtle biases contained in an extraction of model intrinsic scatter, and provide a route to properly account for a non-uniformly sampled data. We also illustrate the implementation of the 2D-fitting problem in hierarchical Bayesian inference software, with an illustrative model using Just Another Gibbs Sampler (\JAGS\footnote{mcmc-jags.sourceforge.net/}).


\section{Mathematical derivation}
\label{section_deriv}

This section develops the general theory for fitting data points in $D$ dimensions by a $(D\!-\!1)$-dimensional plane, such as fitting points in two dimensions by a line (Figure~\ref{fig_schema}). As described in Section \ref{section_introduction} we assume that the data represent a random sample of a population exactly described by a $(D\!-\!1)$-dimensional plane with intrinsic Gaussian scatter, referred to as the `generative model'. The objective is to determine the most likely generative model, by fitting simultaneously for the $(D\!-\!1)$-dimensional plane and the intrinsic Gaussian scatter, given data points with multivariate Gaussian uncertainties: the uncertainties are (1) different for each data point (heteroscedasticity), (2) independent between different data points, and (3) covariant between orthogonal directions, e.g.~$x$-errors and $y$-errors can be correlated. We first present the general method in $D$ dimensions in coordinate-free form (\S\ref{subsection_generaluniform}) and in Cartesian coordinates (\S\ref{subsection_general_cartesian}). Explicit equations for the two-dimensional case are given in \S\ref{subsection_twodim}. Subtleties and generalisations are presented in the appendix.

\subsection{Coordinate-free solution in $D$ dimensions}
\label{subsection_generaluniform}

Consider $N$ data points $i=1,...,N$ in $D$ dimensions. These points are specified by the measured positions $\x_i\in\RD$ relative to a fixed origin $\O\in\RD$ and Gaussian measurement uncertainties expressed by the symmetric covariance matrices $\C_i$. We can then fully express our knowledge about a point $i$ as the probability density function (PDF) $\rho(\x |\x_i)$, describing the probability that point $i$ is truly at the position $\x$ given the measured position $\x_i$. In the case where the intrinsic distribution of points in any direction is uniform (the case of non-uniform point distributions is discussed in Appendix~\ref{subsection_selection}), this PDF is symmetric, i.e.~$\rho(\x|\x_i)=\rho(\x_i|\x)$, and reads
\be\label{eq_rhoi}
	\rho(\x_i|\x) = \frac{1}{\sqrt{(2\pi)^D|\C_i|}}\,e^{-\frac{1}{2}(\x-\x_i)\t\C_i^{-1}(\x-\x_i)}.
\ee
In Figure~\ref{fig_schema}, these PDFs are shown as red shading.

The data points are assumed to be randomly sampled from a PDF, called the generative model, defined by a $(D\!-\!1)$-dimensional plane $\H\subset\RD$ (a line if $D=2$, a plane if $D=3$, a hyperplane if $D\geq4$) with Gaussian scatter $\sigma_\perp$ orthogonal to $\H$. Note that the scatter remains Gaussian along any other direction, not parallel to $\H$. The blue shading in Figure~\ref{fig_schema} represents this generative model, with the solid line indicating the central line $\H$ and the dashed lines being the parallel lines at a distance $\sigma_\perp$ from $\H$. The plane $\H$ is fully specified by the normal vector $\n\perp\H$ going from $\O$ to $\H$, and the distance of a point $\x\in\RD$ from $\H$ equals $|\hat{\n}\t\x-n|$, where $n=|\n|$ and $\hat{\n}=\n/n$. Thus, $\H$ is the ensemble of points $\x$ that satisfy
\be
	\hat{\n}\t\x-n = 0\,.
\ee
Given this parametrisation, the generative model is fully described by the PDF
\be
	\rho_m(\x) = \frac{1}{\sqrt{2\pi\sigma_\perp^2}}\,e^{-\frac{(\hat{\n}\t\x-n)^2}{2\sigma_\perp^2}},
	\label{eq_rhom}
\ee
expressing the probability of generating a point at position $\x$. Note that we here assume that the generative model is infinitely uniform along any dimension in the plane. If the selection function was, e.g., a power law distribution, then $\rho(\x|\x_i) \ne \rho(\x_i|\x)$, and Equation~\ref{eq_rhom} cannot be a simply stated. Such scenarios are discussed in Appendix~\ref{subsection_selection}.

The likelihood of measuring data point $i$ at position $\x_i$ then equals
\be
	\L_i = \int_\RD\!\!d\x\,\rho(\x_i|\x)\rho_m(\x) = \frac{1}{\sqrt{2\pi s_{\perp,i}^2}}\,e^{-\frac{(\hat{\n}\t\x_i-n)^2}{2s_{\perp,i}^2}},
	\label{eq_simpleLi}
\ee
where $s_{\perp,i}^2\equiv\sigma_\perp^2+\hat{\n}\t\C_i\hat{\n}$. Following the Bayesian theorem and assuming uniform priors on $\hat{\n}$ and $\sigma_\perp$, the most likely model maximises the total likelihood $\L=\prod_{i=1}^{N}\L_i$ and hence its logarithm $\ln\L$. Up to an additive constant $\frac{N}{2}\ln(2\pi)$,
\be\
	\ln\L = -\frac{1}{2}\sum_{i=1}^{N} \left[\ln(s_{\perp,i}^2)+\frac{(\hat{\n}\t\x_i-n)^2}{s_{\perp,i}^2}\right].
	\label{eq_likelihood_coordfree}
\ee
Note that Eq.~(\ref{eq_likelihood_coordfree}) is manifestly rotationally invariant in this coordinate-free notation, as expected when fitting data without a preferred choice of predictor / response variables. For the purpose of numerical optimisation, free parameters in the form of unit vectors are rather impractical, because of their requirement to have a fixed norm. It is therefore helpful to express Eq.~(\ref{eq_likelihood_coordfree}) entirely in terms of $\n$, without using $\hat{\n}$ and $n$,
\be
	\ln\L = \!-\frac{1}{2}\sum_{i=1}^{N}\!\left[\ln\!\left(\!\sigma_\perp^2\!\!+\!\frac{\n\!\t\!\C_i\n}{\n\!\t\n}\!\right)\!+\!\frac{(\n\t[\x_i-\n])^2}{\sigma_\perp^2\n\!\t\!\n\!+\!\n\!\t\!\C_i\n}\right]\!.\!\!
	\label{eq_likelihood_coordfree_beta}
\ee

In summary, assuming that the data samples a linear generative model described by Eq.~(\ref{eq_rhom}), the model-parameters that best describe the data are found by maximising the likelihood function given in Eq.~(\ref{eq_likelihood_coordfree}) or Eq.~(\ref{eq_likelihood_coordfree_beta}).

\subsection{Notations in Cartesian coordinates}
\label{subsection_general_cartesian}

For plotting and further manipulation it is often useful to work in a Cartesian coordinate system with $D$ orthogonal coordinates $j=1,...,D$. The typical expression of a linear model in such coordinates is
\be
	x_D = {\Sigma_{j=1}^{D-1}}\alpha_j x_j+\beta,
	\label{eq_hyperplane_coordinates}
\ee

where $x_j$ is the $j$-th coordinate of $\x=(x_1,...,x_D)\t$, $\alpha_j=\partial x_D/\partial x_j$ is the slope along the coordinate $j$, and $\beta$ is the zero-point along the coordinate $D$. Eq.~(\ref{eq_hyperplane_coordinates}) can be rewritten as $\a\t\x+\beta=0$, where $\a\t\equiv(\alpha_1,...,\alpha_{N-1},-1)$ is a normal vector of the $(D\!-\!1)$-dimensional plane. The Gaussian scatter about this plane is now parametrised with the scatter $\sigma_D$ along the coordinate $D$. A vector calculation shows that the transformation between the coordinate-free model parameters $\{\n,\sigma_\perp\}$ and the coordinate-dependent parameters $\{\a,\beta,\sigma_D\}$ takes the form
\be
\begin{split}
	\n & \,=\, -\beta\a(\a\t\a)^{-1}\\
	\sigma_\perp & \,=\, \sigma_D(\a\t\a)^{-1/2}
\end{split}
\ee
or, inversely,
\be
\begin{split}
	\a & \,=\, -\n/n_D \\
	\beta & \,=\, (\n\t\n)/n_D \\
	\sigma_D & \,=\, \sigma_\perp(\n\t\n)^{1/2}/|n_D|.
\end{split}
\label{eq_transformation}
\ee

In practice, one can always fit for $\{\n,\sigma_\perp\}$ by maximising Eq.~(\ref{eq_likelihood_coordfree_beta}) and then convert $\{\n,\sigma_\perp\}$ to $\{\a,\beta,\sigma_D\}$, if desired. Alternatively, we can rewrite Eq.~(\ref{eq_likelihood_coordfree_beta}) directly in terms of the parameters $\{\a,\beta,\sigma_D\}$ as
\be
	\ln\L = \frac{1}{2}\sum_{i=1}^{N} \left[\ln\frac{\a\t\a}{\sigma_D^2+\a\t\C_i\a}-\frac{(\a\t\x_i+\beta)^2}{\sigma_D^2+\a\t\C_i\a}\right],
	\label{eq_likelihood_coordinates}
\ee
however, this formulation of the likelihood is prone to optimisation problems since the parameters diverge when $\H$ becomes parallel to the coordinate $D$.
\subsection{Two-dimensional case}
\label{subsection_twodim}

Let us now consider the special case of $N$ points in $D=2$ dimensions (Figure~\ref{fig_schema}), adopting the standard `$xy$' notation, where $x$ replaces $x_1$ and $y$ replaces $x_2=x_D$. The data points are specified by the positions $\x_i=(x_i,y_i)\t$ and heteroscedastic Gaussian errors $\sxi$ and $\syi$ with correlation coefficients $\ci$; thus the covariance matrix elements are $(\C_i)_{xx}=\sxi^2$, $(\C_i)_{yy}=\syi^2$, $(\C_i)_{xy}=\sxyi=\ci\sxi\syi$. These points are to be fitted by a line
\be
	y = \alpha x+\beta,
\ee
which we can also parameterise by the normal vector $\n=n(\cos\phi,\sin\phi)\t$ going from the origin to the line. The parameters $\alpha$, $\beta$, and $\sigma_y$ ($\sigma_y=\sigma_D$ being the intrinsic scatter along $y$) are then given by (following Eqs.~\ref{eq_transformation})
\be
	\alpha = \frac{-1}{\tan\phi},~~~\beta=\frac{n}{\sin\phi},~~~\sigma_y=\frac{\sigma_\perp}{|\sin\phi|},
\ee
which manifestly diverge as the line becomes vertical ($\phi\in\pi\Z$). In these notations, Eqs.~(\ref{eq_likelihood_coordfree}) and (\ref{eq_likelihood_coordinates}) simplify to
\be
\begin{split}
	\!\ln\L & = -\frac{1}{2}\sum_{i=1}^{N} \left[\ln s^2_{\perp,i}\!+\frac{(x_i\cos\phi\!+\!y_i\sin\phi\!-\!n)^2}{s^2_{\perp,i}}\right]\\
	& = \frac{1}{2}\sum_{i=1}^{N} \left[\ln\frac{\alpha^2+1}{s^2_{y,i}}-\frac{(\alpha x_i-y_i+\beta)^2}{s^2_{y,i}}\right],
\end{split}
\label{eq_likelihood2D}
\ee
where $s^2_{\perp,i}\equiv\sigma_\perp^2+\sxi^2\cos^2\!\phi+\syi^2\sin^2\!\phi+2\sxyi\cos\phi\sin\phi$ and $s^2_{y,i}\equiv\sigma_y^2+\sxi^2\alpha^2+\syi^2-2\sxyi\alpha$ (reminder: $\sigma_y$ is the intrinsic vertical scatter of the model, whereas $\syi$ is the vertical uncertainty of point $i$). The most likely fit to the data is obtained by maximising Eq.~(\ref{eq_likelihood2D}), either through adjusting the parameters $(\phi,n,\sigma_\perp)$ or $(\alpha,\beta,\sigma_y)$. Following Appendix~\ref{subsection_bias}, \emph{unbiased estimators for the population-model (as opposed to the sample-model)} of the scatter $\sigma_\perp$ or $\sigma_y$ and its variance $\sigma_\perp^2$ or $\sigma_y^2$ can be obtained via Eqs.~(\ref{eq_bias2}) and (\ref{eq_bias1}), respectively.

It is worth stressing that the exact Eq.~(\ref{eq_likelihood2D}) never reduces to the frequently used, simple minimisation of $\chi^2\equiv\sum_i(y_i-\alpha x_i-\beta)^2$, not even if the data has zero uncertainties and if the intrinsic scatter is fixed. This reflects the fact that the $\chi^2$-minimisation breaks the rotational invariance by assuming a uniform prior for the slope $\alpha$ (rather than the angle $\phi$). The term $\ln(\alpha^2+1)$ in Eq.~(\ref{eq_likelihood2D}) is needed to recover the rotational invariance. 

\section{Numerical implementation}
\label{section_hyper.fit}
Having established the above mathematical framework for calculating the likelihood for a multi-dimensional fit, we developed a fully documented and tested \R package called \hf. This is available through a github repository\footnote{https://github.com/asgr/hyper.fit} and can be installed from within \R with the following commands\\

\noindent {\tt
> install.packages("magicaxis", dependencies=TRUE)\\
> install.packages("MASS", dependencies=TRUE)\\
> install.packages("rgl", dependencies=TRUE)\\
> install.packages("devtools")\\
> library(devtools)\\
> install\_github("asgr/LaplacesDemon")\\
> install\_github("asgr/hyper.fit")\\
}

To load the package into your \R session run\newline

\noindent {\tt
> library(hyper.fit)
}\newline

The core function that utilises the maths presented in the previous Section is {\tt hyper.like}, which calculates the likelihood (Eq.~(\ref{eq_likelihood_coordfree_beta})) for a given set of hyperplane parameters and a given dataset. This dataset can be error-free, or include covariant and heteroscedastic errors. To fit a hyperplane the {\tt hyper.like} function is accessed by {\tt hyper.fit}, a utility fitting function that attempts to optimally fit a generative model to the data. The package includes a number of user-friendly summary and plotting outputs in order to visualise the fitted model and investigate its quality. The full 35 page manual detailing the \hf package can be viewed at hyperfit.icrar.org, and is included within the help files of the package itself (e.g. see {\tt > ?hyper.fit} within \R once the package is loaded.).

\subsection{{\tt hyper.like}}\label{subsection_hyperlike}

The basic likelihood function {\tt hyper.like} requires the user to specify a vector of model parameters {\tt parm}, an $N\times D$ dimensional position matrix {\tt X} (where $N$ is the number of data points and $D$ the dimensionality of the dataset, i.e.\ 10 rows by 2 columns in the case of 10 data points with $x$ and $y$ positions) and a $D \times D \times N$ array {\tt covarray} containing the covariance error matrix for each point stacked along the last index. A simple example looks like
\newline

\noindent {\tt
> hyper.like(parm, X, covarray)
}
\newline

\noindent {\tt parm} must be specified as the normal vector $\n$ that points from the origin to the hyperplane concatenated with the intrinsic scatter $\sigma_\perp$ orthogonal to the hyperplane. E.g.\ if the user wants to calculate the likelihood of a plane in 3D with equation $z=2x+3y+1$ and intrinsic scatter along the $z$ axis of 4, this becomes a normal vector with elements $[-0.143, -0.214, 0.071]$ with intrinsic scatter orthogonal to the hyperplane equal to 1.07. In this case {\tt parm=c(-0.143, -0.214, 0.071, 1.07)}, i.e. as expected, a 3D plane with intrinsic scatter is fully described by 4~$(=D+1)$ parameters. Because of the large number of coordinate systems that unambiguously define a hyperplane (and many of them might be useful in different circumstances) the \hf package also includes the function {\tt hyper.convert} that allows the user to convert between coordinate systems simply. In astronomy the Euclidian $z=\alpha[1]x+\alpha[2]y+\beta$ system is probably the most common \citep[e.g.\ Fundamental Plane definition][]{fabe87,binn98}, but this has limitations when it comes to efficient high dimensional hyperplane fitting, as we discuss later.

\subsection{{\tt hyper.fit}}

In practice, we expect most user interaction with the \hf package will be via the higher level {\tt hyper.fit} function (hence the name of the package). For example, to find an optimal fit for the data in the $N\times D$ dimensional matrix {\tt X} (see \S\ref{subsection_hyperlike}) with no specified error, it suffices to run
\newline

\noindent {\tt
> fit=hyper.fit(X)}
\newline

This function interacts with the {\tt hyper.like} function and attempts to find the best generative hyperplane model via a number of schemes. The user has access to three main high level fitting routines: the base \R~{\tt optim} function (available with all installations), the {\tt LaplaceApproximation} function and the {\tt LaplacesDemon} function (both available in the \ld package).

\subsubsection{{\tt optim}}

{\tt optim} is the base \R parameter optimisation routine. It allows the user to generically find the maximum or minimum of a target function via a number of built-in schemes that are popular in the statistical optimisation community. The default option is to use a Nelder-Mead (Nelder \& Mead 1965) optimiser that uses only function values. This is robust but relatively slow, with the advantage that it will work reasonably well for non-differentiable functions, i.e.\ it does not compute the local hessian at every step. Conjugate gradient \citep{hest52}; quasi-Newtonian BFGS \citep[Broyden, Fletcher, Goldfarb and Shanno;][]{flet70}; and simulated annealing \citep{beli92} are some of the other popular methods available to the user. All of these schemes are fully documented within the {\tt optim} manual pages that is included with any standard \R implementation.

\subsubsection{{\tt LaplaceApproximation}}

To offer the user more flexibility we also allow access to the non-base open-source MIT-licensed \ld package that users can install from github.com/asgr/LaplacesDemon (the direct \R installation code is provided above). It is developed by Statisticat (LLC) and currently contains 18 different optimisation schemes. Many of these are shared with {\tt optim} (e.g.\ the default option of Nelder-Mead; conjugate gradient descent), but many are unique to the package (e.g.\ Levenberg-Marquardt, Marquardt 1963; particle swarm optimisation, Kennedy \& Eberhart 1995).

The {\tt LaplaceApproximation} function offers a host of high level user tools, and with these object oriented extensions it tends to be slower to converge compared to the same method in {\tt optim}. The large range of methods means the user should be able to find a robustly converged solution in most situations, without having to resort to much more computationally expensive MCMC methods accessed through the {\tt LaplacesDemon} function which is also available in the \ld package.

The {\tt LaplaceApproximation} function and the wider \ld package is extensively documented, both in the bundled manual and at the www.bayesian-inference.com website \footnote{https://web.archive.org/web/20141224051720/ http://www.bayesian-inference.com/index} maintained by the package developers.

\subsubsection{{\tt LaplacesDemon}}

The final hyperplane fitting function that users of \hf have access to is {\tt LaplacesDemon} which is also included in the \ld package. The function includes a suite of 41 Markov-Chain Monte-Carlo (MCMC) routines, and the number is still growing. For \hf, the default routine is Griddy-Gibbs \citep{ritt92}, a component-wise algorithm that approximates the conditional distribution by evaluating the likelihood model at a set of points on a grid. This proved to have attractive qualities and well behaved systematics for relatively few effective samples during testing on toy models. The user also has access to a large variety of routine families, including Gibbs; Metropolis-Hasting \citep{hast70}; slice sampling \citep{neal03}; and Hamiltonian Monte Carlo \citep{duan87}.

In practice the user should investigate the range of options available, and the optimal routine for a given problem might not simply be the default provided. To aid the user in choosing an appropriate scheme, the \ld package includes a comprehensive suite of analysis tools to analyse the post convergence Markov-Chains. The potential advantages to using an MCMC approach to analyse the generative likelihood model are manifold, allowing the user of \hf to investigate complex high-dimensional multi-modal problems that might not be well-suited to the traditional downhill gradient approaches included in {\tt optim} and {\tt LaplaceApproximation}. Again, the function is extensively described both in the \ld package and on the www.bayesian-inference.com website \footnote{https://web.archive.org/web/20141224051720/ http://www.bayesian-inference.com/index}.

\subsection{{\tt plot}}

The \hf package comes with a class sensitive plotting function where the user is able to execute a command such as\\

\noindent {\tt
> plot(hyper.fit(X))\\
}\newline
where X is a $2\times N$ or $3\times N$ matrix. The plot displays the data with the best fit. If (uncorrelated or correlated) Gaussian errors are given, e.g.~via {\tt plot(hyper.fit(X,covarray))}, they are displayed as 2D ellipses or 3D ellipsoids, respectively. The built-in plotting only works in 2 or 3 dimensions (i.e.\ x,y or x,y,z data), which will likely be the two most common hyperplane fitting situations encountered in astronomical applications.

\subsection{{\tt summary}}

The \hf package comes with a class sensitive summary function where the user is able to execute a command such as\\

\noindent{\tt
> summary(hyper.fit(X))\\
}\newline
where X can be any $(D\times N)$-array with $D\geq2$ dimensions. Using {\tt summary} for 2D data will produce an output similar to\\

\noindent{\small\tt
> summary(fitnoerror)\\
Call used was:\\
hyper.fit(X = cbind(xval, yval))\\
\\
Data supplied was 5rows $\times$ 2cols.\
\\
Requested parameters:\\
 alpha1 beta.vert scat.vert \\
0.4680861 0.6272718 0.2656171 \\
\\
Errors for parameters:\\
 err\_alpha1 err\_beta.vert err\_scat.vert \\
 0.12634307 0.11998805 0.08497509 \\
 \\
The full parameter covariance matrix:\\
    alpha1  beta.vert  scat.vert\\
alpha1  0.015962572 -0.0021390417 0.0016270601\\
beta.vert -0.002139042 0.0143971319 -0.0002182796\\
scat.vert 0.001627060 -0.0002182796 0.0072207660\\
\\
The sum of the log-likelihood for the best fit parameters:\\
  LL \\
4.623924 \\
\\
Unbiased population estimator for the intrinsic scatter:\\
scat.vert \\
0.3721954 \\
\\
Standardised parameters vertically projected along dimension 2 (yval):\\
 alpha1 beta.vert scat.vert \\
0.4680861 0.6272718 0.2656171 \\
\\
Standardised generative hyperplane equation with unbiased population estimator for the intrinsic scatter, vertically projected along dimension 2 (yval):\\
yval $\sim$ N(mu= 0.4681*xval + 0.6273, sigma= 0.3722)\\
}

\subsection{Web interface}

\begin{figure}
	\centering
	\includegraphics[width=\columnwidth]{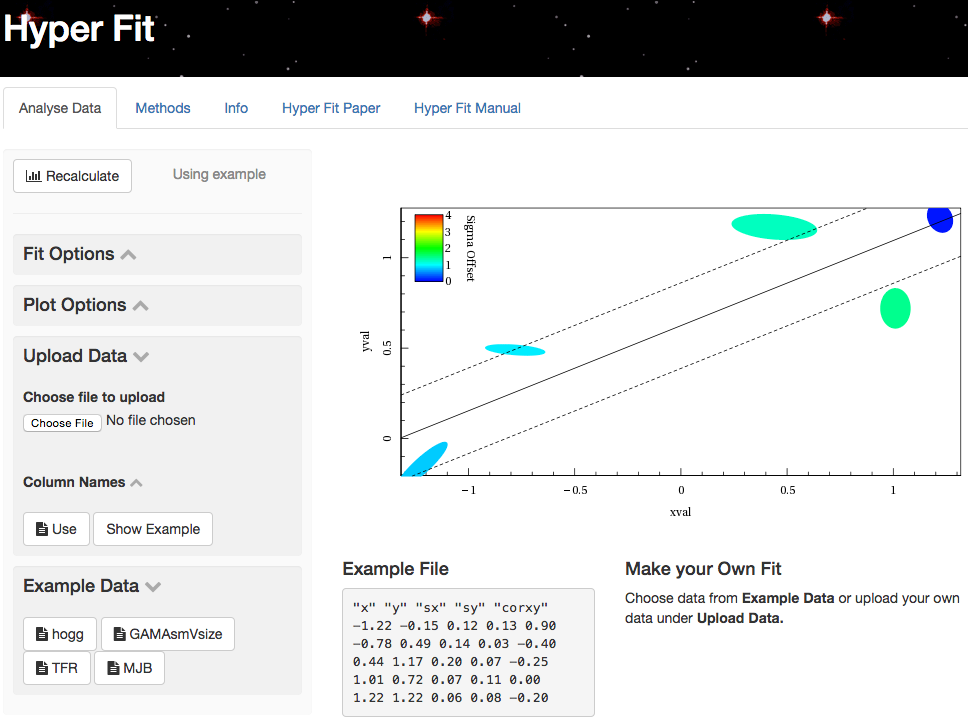}
	\caption{The front page view of the \hf web tool available at hyperfit.icrar.org. The web tool allows users to interact with \hf through a simple GUI interface, with nearly all \hf functionality available to them. The code itself runs remotely on a machine located at ICRAR, and the user's computer is only used to render the web site graphics.
	\label{fig_webtool}}
\end{figure}

To aid the accessibility of the tools we have developed in this paper, a high level web interface to the \hf package has been developed that runs on a permanent virtual server using \R~\Shiny\footnote{http://shiny.rstudio.com/}. The \hf computations are executed remotely, and the web interface itself is in standard JavaScript and accessible using any modern popular web browser. This website interface to \hf can be accessed at hyperfit.icrar.org. The \hf website also hosts the latest version of the manual, and the most recent version of the associated \hf paper. The design philosophy for the page is heavily influenced by modern web design, where the emphasis is on clean aesthetics, simplicity and intuitiveness. The front page view of the web interface is shown in Figure \ref{fig_webtool}.

The \hf web tool allows access to nearly all of the functionality of \hf, with a few key restrictions:

\begin{itemize}
\item Only 2D or 3D analysis available (\R \hf allows for arbitrary $N$D analysis)
\item Uploaded files are limited to 2,000 or fewer row entries (\R \hf allows for hard-disk limited size datasets)
\item When using the {\tt LaplaceApproximation} or {\tt LaplacesDemon} algorithms only 20,000 iterations are allowed (\R \hf has no specific limit on the number of iterations)
\item A number of the more complex {\tt LaplacesDemon} methods are unavailable to the user because the specification options are too complex to describe via a web interface (\R \hf allows access to all {\tt LaplacesDemon} methods)
\end{itemize}

In practice, for the vast majority of typical use cases these restrictions will not limit the utility of the analysis offered by the web tool. For the extreme combination of 2,000 rows of data and 20,000 iterations of one of the {\tt LaplacesDemon} MCMC samplers users might need to wait a couple of minutes for results.

\section{Mock examples}
\label{section_mockexamples}
In this Section we give some examples using two sets of idealised mock data.

\subsection{Simple Examples}

To help novice users get familiar with \R and the \hf package we start this Section by building the basic example shown in Figure~\ref{fig_schema}. We first create the central $x$ and $y$ positions of the 5 data points, encoded in the 5-element vectors {\tt x\_val} and {\tt y\_val},\\

\noindent {\tt
> xval = c(-1.22, -0.78, 0.44, 1.01, 1.22)\\
> yval = c(-0.15, 0.49, 1.17, 0.72, 1.22)\\
}

\begin{figure}
	\centering
	\includegraphics[width=\columnwidth]{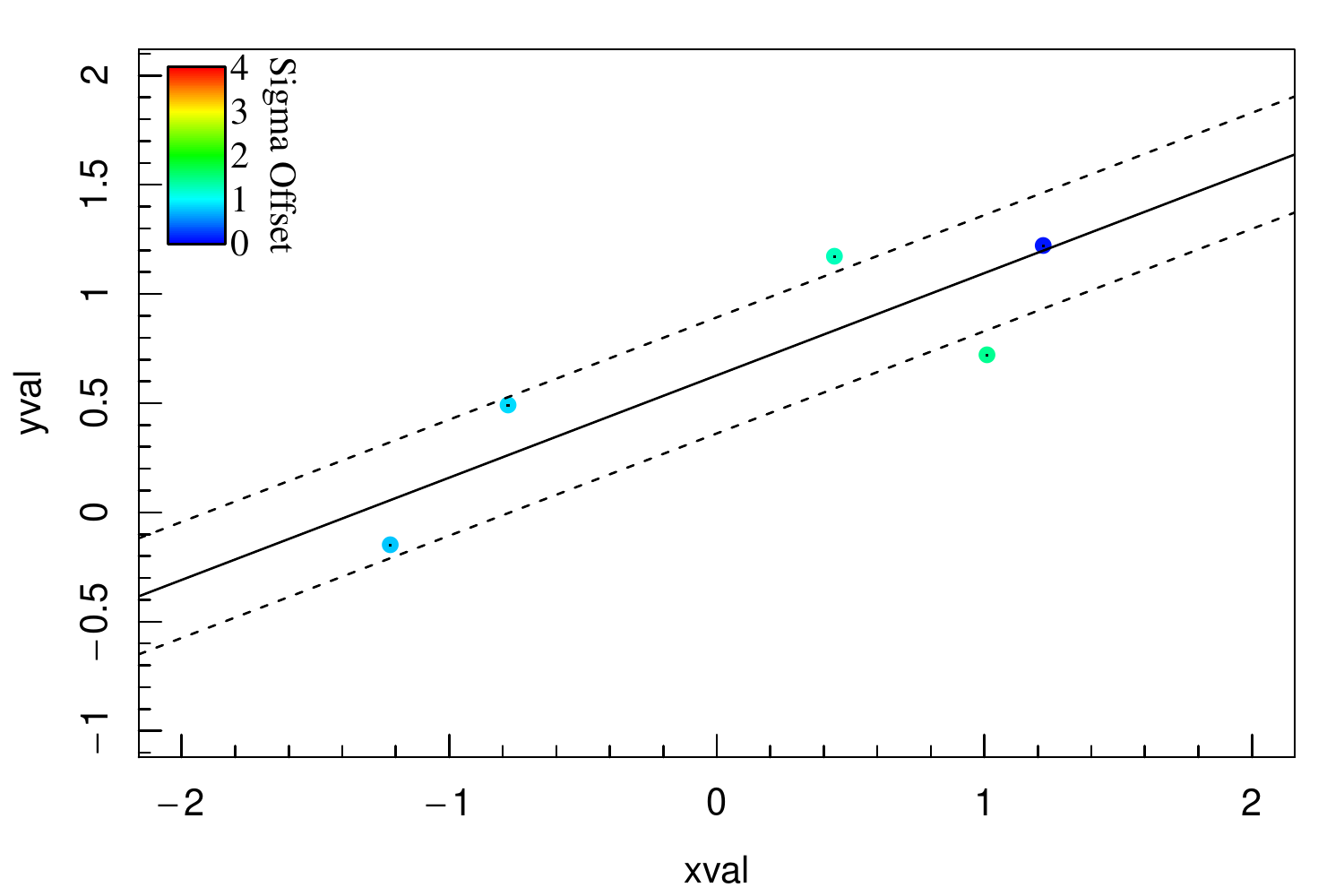}
	\caption{2D fit with no errors. Figure shows the default plot output of the {\tt hyper.plot2d} function (accessed via the class specific {\tt plot} method) included as part of the \R \hf package, where the best generative model for the data is shown as a solid line with the intrinsic scatter indicated by dashed lines. The colouring of the points shows the `tension' with respect to the best-fit linear model and the measurement errors (zero in this case), where redder colours indicate data that is less likely to be explainable.
	\label{fig_fitnoerror}}
\end{figure}

We can can obtain the \hf solution and plot the result (seen in Figure \ref{fig_fitnoerror}) with\\

\noindent{\tt
> fitnoerror=hyper.fit(cbind(xval, yval))\\
> plot(fitnoerror) \hfill \textrm{(Figure \ref{fig_fitnoerror})}
}\newline

\begin{figure}
	\centering
	\includegraphics[width=\columnwidth]{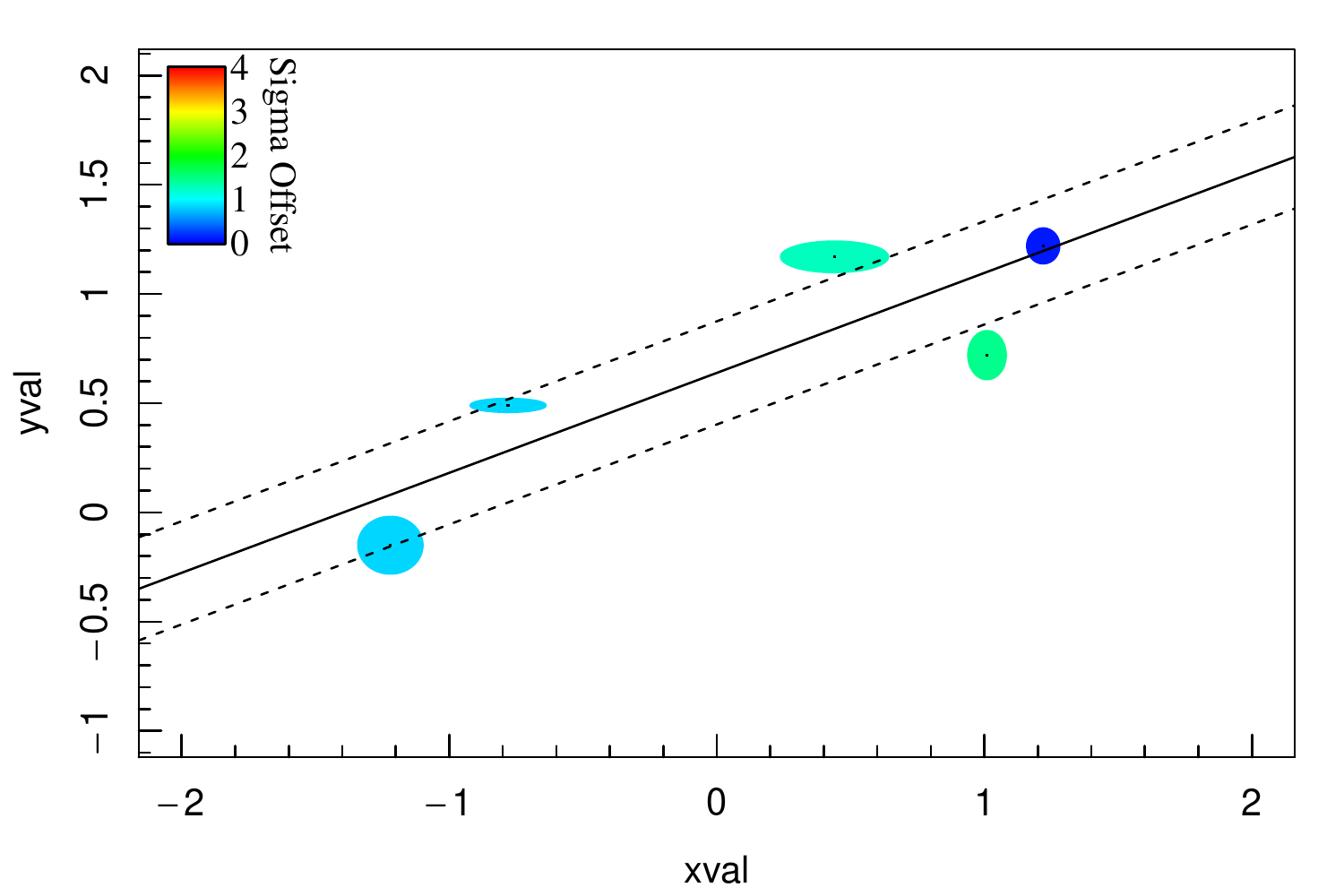}
	\caption{2D fit with uncorrelated (between $x$ and $y$) errors. The errors are represented by 2D ellipses at the location of the $xy$-data. See Figure \ref{fig_fitnoerror} for further details on this Figure. The reduction in the intrinsic scatter required to explain the data compared to Figure \ref{fig_fitnoerror} is noticeable (see the dashed line intersects on the y-axis).
	\label{fig_fitwitherror}}
\end{figure}

Since this example has no errors associated with the data positions the generative model created has a large intrinsic scatter that broadly encompasses the data points (dashed lines in Figure \ref{fig_fitnoerror}). Extending the analysis we can add $x$-errors {\tt x\_err} (standard deviations of the errors along the $x$-dimension) and $y$-errors {\tt y\_err} and recalculate the fit and plot (Figure \ref{fig_fitwitherror}) with\\

\noindent {\tt
> xerr = c(0.12, 0.14, 0.20, 0.07, 0.06)\\
> yerr = c(0.13, 0.03, 0.07, 0.11, 0.08)\\
> fitwitherror=hyper.fit(cbind(xval, yval), vars=cbind(xerr, yerr)\^{}2)\\
> plot(fitwitherror) \hfill \textrm{(Figure \ref{fig_fitwitherror})}
}\newline

\begin{figure}
	\centering
	\includegraphics[width=\columnwidth]{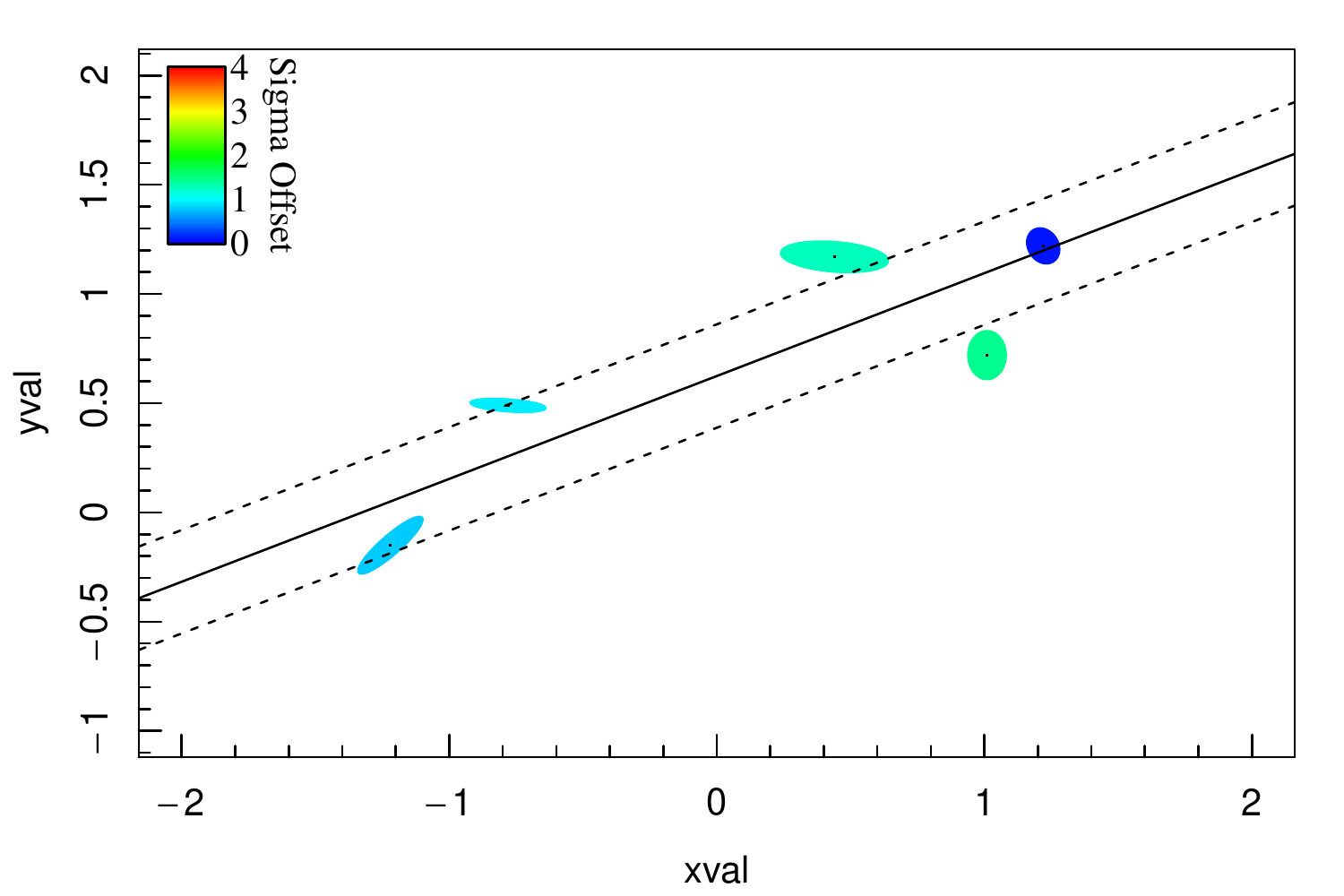}
	\caption{2D fit with correlated (between $x$ and $y$) errors. The errors are represented by 2D ellipses at the location of the $xy$-data. See Figure \ref{fig_fitnoerror} for further details on this Figure. The keen reader might notice that this Figure uses data from the Figure \ref{fig_schema} schematic.
	\label{fitwitherrorandcor}}
\end{figure}

Since we have added errors to the data, the intrinsic scatter required in the generative model is reduced. To extend this 2D example to the full complexity of Figure~\ref{fig_schema} we assume that we know {\tt xy\_cor} (the correlation between the errors for each individual data point) and recalculate the fit and plot (Figure \ref{fitwitherrorandcor}) with\\

\noindent {\tt
> xycor = c(0.90, -0.40, -0.25, 0.00, -0.20)\\
> fitwitherrorandcor=hyper.fit(cbind(xval, yval), covarray=makecovarray2d(xerr, yerr, xycor))\\
> plot(fitwitherrorandcor) \hfill \textrm{(Figure \ref{fitwitherrorandcor})}
}\newline

\begin{figure}
	\centering
	\includegraphics[width=\columnwidth]{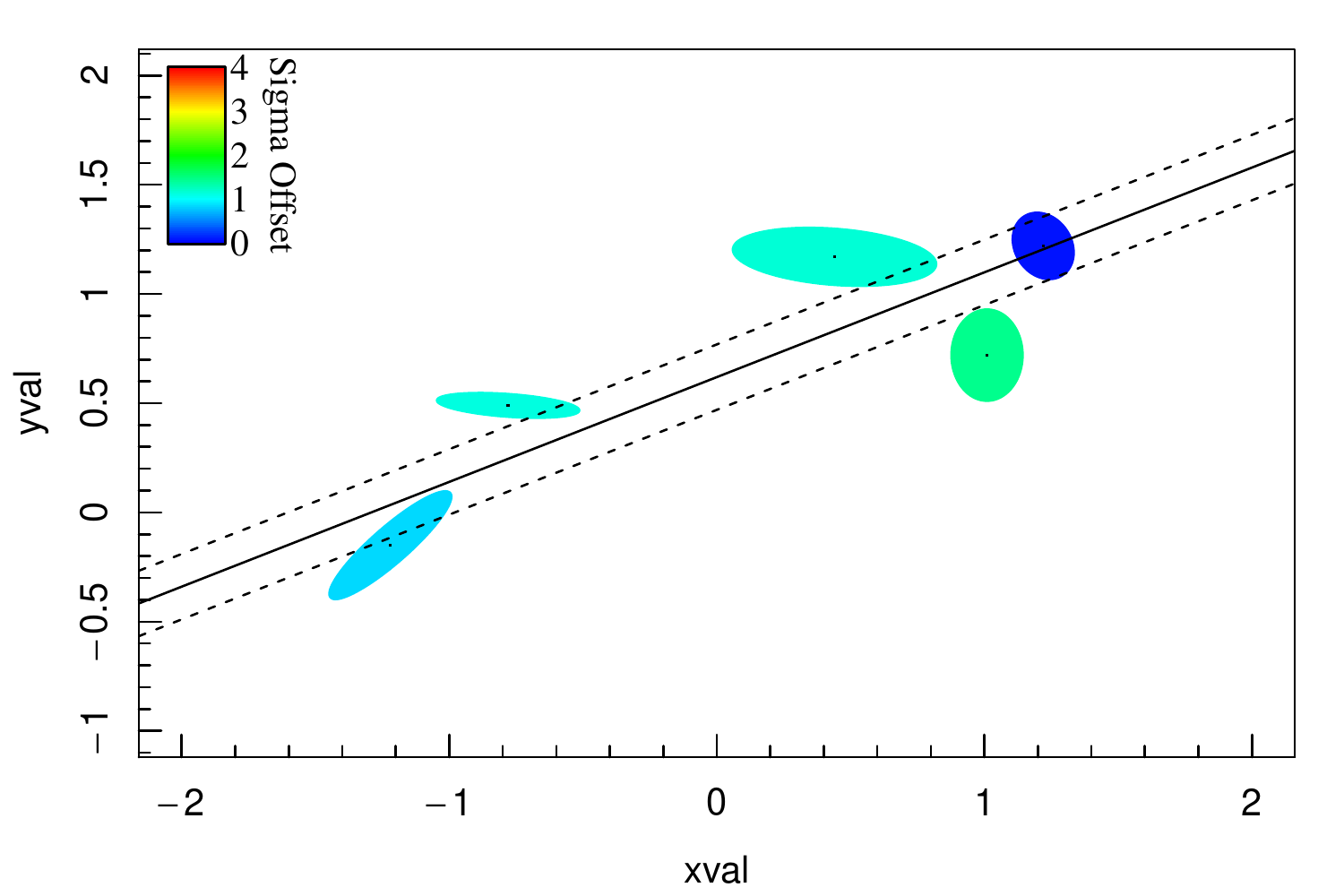}
	\caption{2D fit with correlated (between $x$ and $y$) errors. The errors here are a factor 1.9 times larger than used in \ref{fitwitherrorandcor}. See Figure \ref{fig_fitnoerror} for further details on this Figure.
	\label{fitwithbigerrorandcor}}
\end{figure}

The effect of including the correlations can be quite subtle, but inspecting the intersections on the $y$-axis it is clear that both the best-fitting line and the intrinsic scatter have slightly changed. To see how dramatic the effect of large errors can be on the estimated intrinsic scatter we can simply inflate our errors by a large factor. We recalculate the fit and plot (Figure \ref{fitwithbigerrorandcor}) with\\

\noindent {\tt
> fitwithbigerrorandcor=hyper.fit(cbind(xval, yval), covarray=makecovarray2d(xerr*1.9, yerr*1.9, xycor))\\
> plot(fitwithbigerrorandcor) \hfill \textrm{(Figure \ref{fitwithbigerrorandcor})}
}\newline

\begin{figure}
	\centering
	\includegraphics[width=\columnwidth]{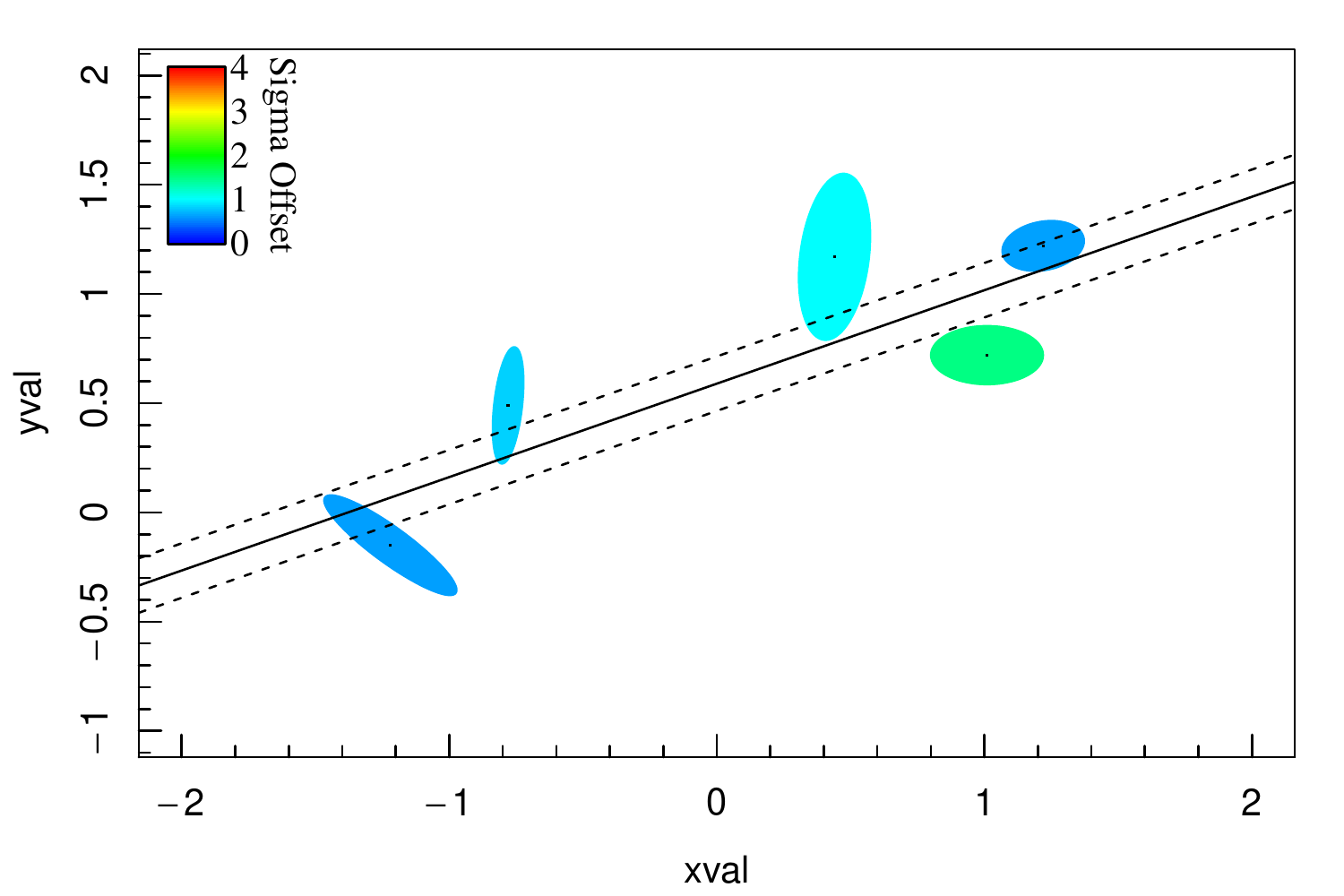}
	\caption{2D fit with correlated (between $x$ and $y$) errors. The errors here are a factor 1.9 times larger than used in \ref{fitwitherrorandcor} and the correlation matrix is rotated by 90 degrees. See Figure \ref{fig_fitnoerror} for further details on this Figure.
	\label{fitwithbigerrorandrotcor}}
\end{figure}

Per-data-point error covariance is often overlooked during regression analysis because of the lack of readily available tools to correctly make use of the information, but the impact of correctly using them can be non-negligible. As a final {\it simple} example we take the input for Figure \ref{fitwithbigerrorandcor} and simply rotate the covariance matrix by 90 degrees. We recalculate the fit and plot (Figure \ref{fitwithbigerrorandrotcor}) with\\

\noindent {\tt
> fitwithbigerrorandrotcor=hyper.fit(cbind(xval, yval), covarray=makecovarray2d(yerr*1.9, xerr*1.9, -xycor))\\
> plot(fitwithbigerrorandrotcor) \hfill \textrm{(Figure \ref{fitwithbigerrorandrotcor})}
}\newline

The impact of simply rotating the covariance matrix is to flatten the preferred slope and to slightly reduce the intrinsic scatter.

%

\subsection{Hogg Example}

\begin{figure}
	\centering
	\includegraphics[width=\columnwidth]{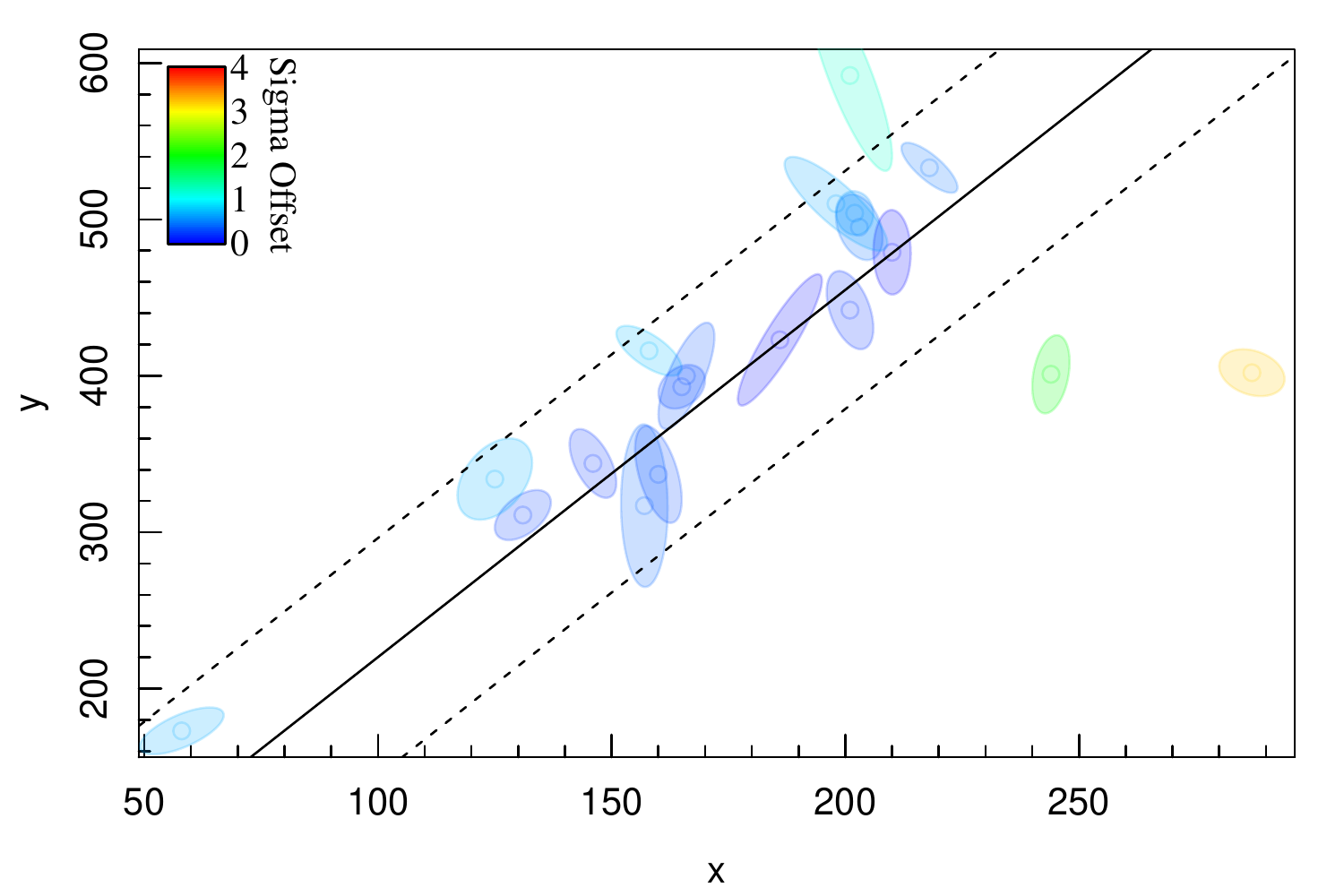}
	\includegraphics[width=\columnwidth]{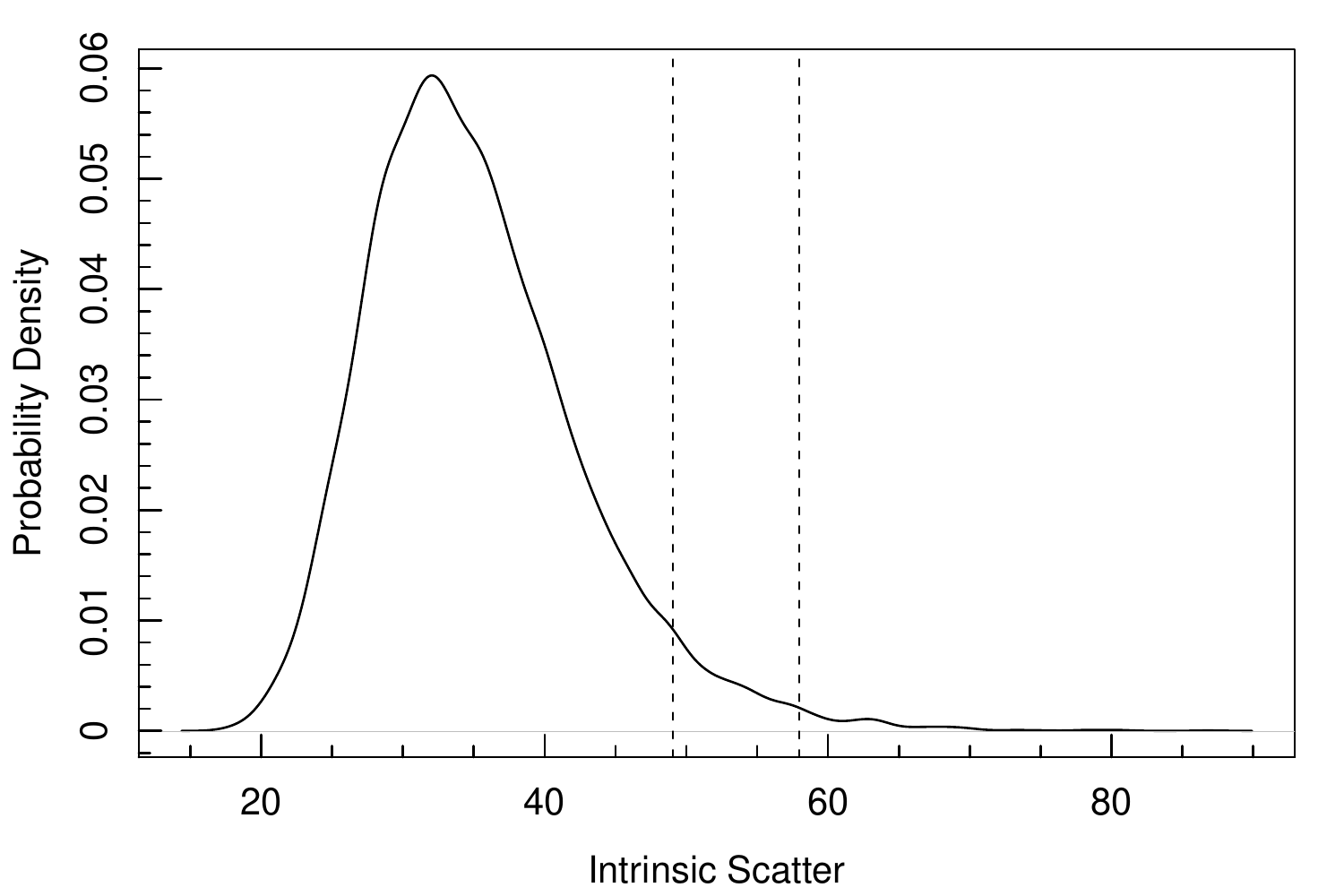}
	\caption{
	2D toy data with correlated (between $x$ and $y$) errors taken from \citet{hogg10}. The top panel shows the fit to the \citet{hogg10} data minus row 3, as per exercise 17 and Figure 13 of \citet{hogg10}, the bottom panel shows the MCMC posterior chains for the intrinsic scatter, as per Figure 14 of \citet{hogg10}. The vertical dashed lines specify the 95$^{th}$ and 99$^{th}$ percentile range of the posterior chains, as requested for the original exercise. See Figure \ref{fig_fitnoerror} for further details on the top two panels of this Figure. 
	\label{fig_hogg}}
\end{figure}

In the arXiv paper by \citet{hogg10} the general approach to the problem of generative fitting is explored, along with exercises for the user. The majority of the paper is concerned with 2D examples, and an idealised data set (Table 1 in the paper) is included to allow readers to recreate their fits and plots. This dataset comes bundled with the \hf package, making it easy to recreate many of the examples of \citet{hogg10}. Towards the end of their paper they include an exercise on fitting 2D data with covariant errors. We can generate the \hf solution to this exercise and create the relevant plots (Figure \ref{fig_hogg}) with\\

\noindent{\tt
> data(hogg)\\
> hoggcovarray= makecovarray2d(hogg[-3,"x\_err"], hogg[-3,"y\_err"], hogg[-3,"corxy"])\\
> plot(hyper.fit(hogg[-3,c("x", "y")], covarray=hoggcovarray)) \hfill \textrm{(Figure \ref{fig_hogg})}
}\newline

The fit we generate agrees very closely with the example in \citet{hogg10}, which is not surprising since the likelihood function we maximise is identical, bar a normalising factor (i.e.~an additive constant in $\ln\mathcal L$). The differences seen between Hogg's Figure 14 and the bottom panel of our Figure \ref{fig_hogg}, showing the projected posterior chain density distribution of the intrinsic scatter parameter, are likely to be due to our specific choice of MCMC solver. In our \hf example we use Componentwise Hit-And-Run Metropolis \citep[CHARM,][]{turc71}, being a particularly simple MCMC for a novice user to specify, and one that performs well as long as the number of data points is greater than $\sim$10.


\section{Astrophysical Examples}
\label{section_astroexamples}
In this Section, we have endeavoured to include a mixture of real examples from published astronomy papers. In some cases aspects of the data are strictly proprietary (in the sense that the tables used to generate paper Figures have not been explicitly published), but the respective authors have allowed us to include the required data in our package. 

\subsection{2D Mass-Size Relation}

\begin{figure}
	\centering
	\includegraphics[width=\columnwidth]{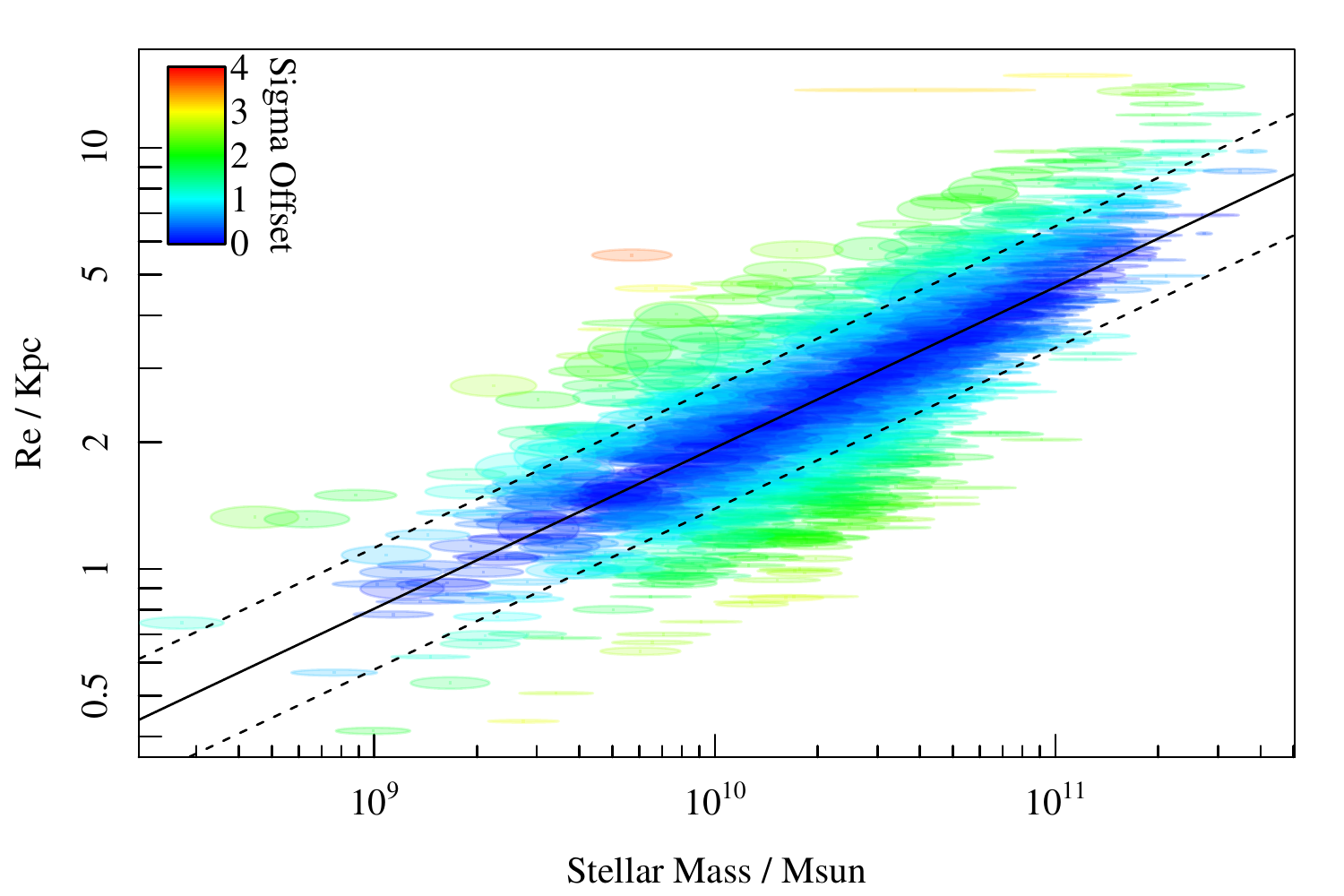}
	\caption{GAMA mass-size relation data taken from Lange et~al in prep. See Figure \ref{fig_fitwitherror} for further details on this Figure.
	\label{fig_MS}}
\end{figure}

The galaxy mass-size relation is a topic of great popularity in the current astronomical literature \citep{shen03, truj04, lang15}. The \hf package includes data taken from the bottom-right panel of Figure 3 from \citet{lang15}. To do our fit on these data and create Figure \ref{fig_MS} we make use of the published data errors and the $1/V_{max}$ weights calculated for each galaxy by running\\

\noindent {\tt
> data(GAMAsmVsize)\\
> plot(hyper.fit(GAMAsmVsize[,c("logmstar", "logrekpc")], vars=GAMAsmVsize[,c("logmstar\_err", "logrekpc\_err")]\^{}2, weights=data[,"weights"])) \hfill \textrm{(Figure \ref{fig_MS})}
}\newline

This analysis produces a mass-size relation of

\ben
\begin{split}
\log_{10}R_e \sim~& \mathcal N[\mu=(0.382\pm0.006) \log_{10} \mathcal M_{*}-(3.53\pm0.06),\\
 &\sigma=0.145\pm0.003],
\end{split}
\een
\noindent where $R_e$ is the effective radius of the galaxy in kpc, $\mathcal N$ is the normal distribution with parameters $\mu$ (mean) and $\sigma$ (standard deviation) and $\mathcal M_{*}$ is the stellar mass in solar mass units. The relation published in \citet{lang15} is $\log_{10}R_e = 0.46\pm0.02 (\log_{10} \mathcal M_{*})-4.38\pm0.03$. The function in \citet{lang15} is noticeably steeper due to being fit to the running median with weighting determined by the spread in the data and the number of data points in the stellar mass bin, rather than directly to the data including the error.

\subsection{2D Tully-Fisher Relation}

\begin{figure}
	\centering
	\includegraphics[width=\columnwidth]{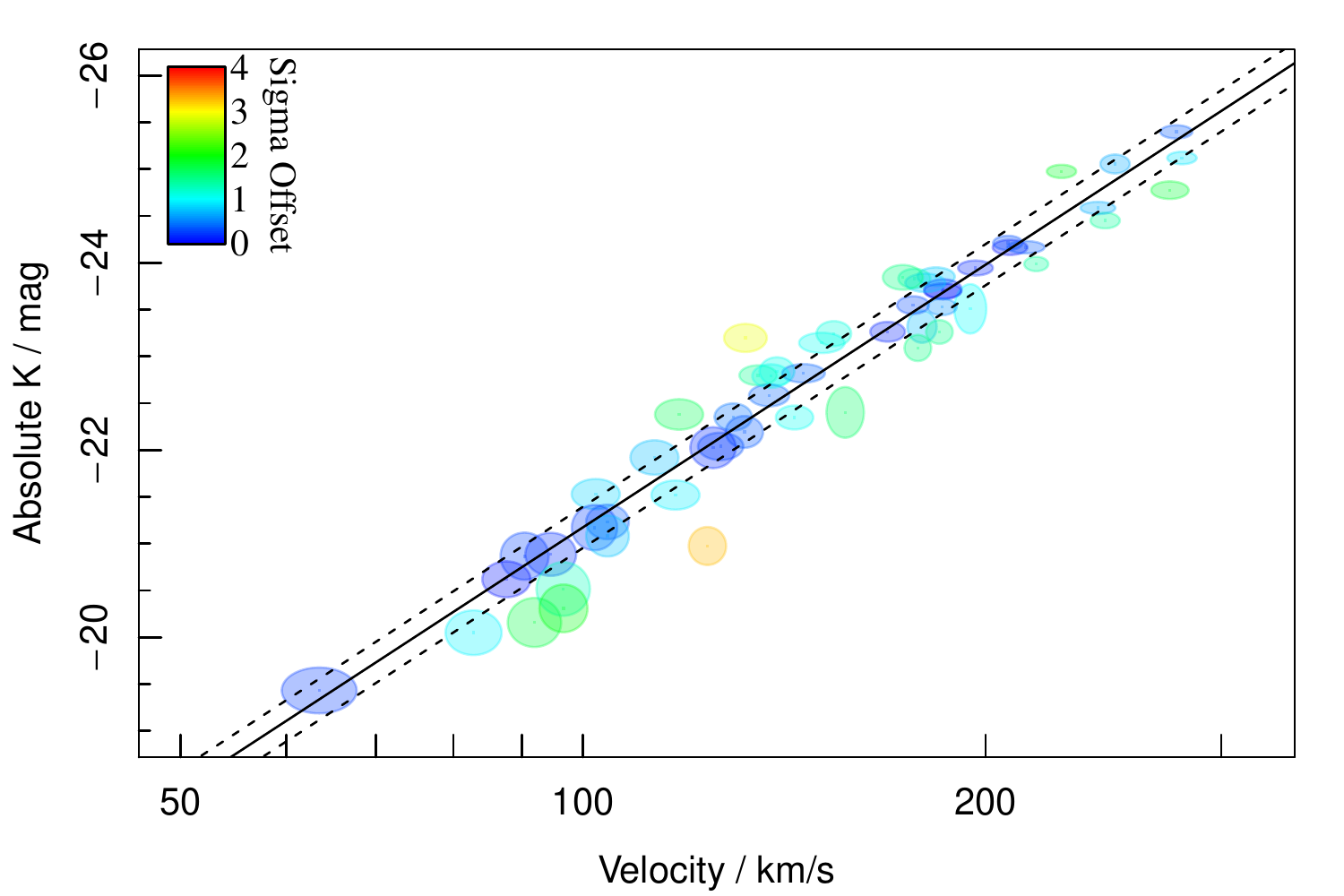}
	\caption{Tully-Fisher data taken from \citet{obre13}, with the best-fit \hf generative model for the data shown as a solid line and the intrinsic scatter indicated with dashed lines. See Figure \ref{fig_fitnoerror} for further details on this Figure.
	\label{fig_TFR}}
\end{figure}

One of the most common relationships in extragalactic astronomy is the so-called Tully-Fisher relation \citep[TFR,][]{tull77}. This tight coupling between inclination corrected disc rotation velocity and the absolute magnitude (or, more fundamentally, stellar mass) of spiral galaxies is an important redshift-independent distance indicator, as well as a probe for galaxy evolution studies. Here we use TFR data included in the \hf package that has been taken from Figure 4 of \citet{obre13}, where we fit the data and create Figure \ref{fig_TFR} with\\

\noindent {\tt
> data(TFR)\\
> plot(hyper.fit(TFR[,c("logv", "M\_K")], vars=TFR[,c("logv\_err", "M\_K\_err")]\^{}2)) \hfill \textrm{(Figure \ref{fig_TFR})}
}\newline

This yields a TFR of

\ben
\begin{split}
M_K \sim~ &\mathcal N[\mu=(-9.3\pm0.4) \log_{10} v_{circ}-(2.5\pm0.9),\\
& \sigma=0.22\pm0.04],
\end{split}
\een

\noindent where $M_K$ is the absolute $K$-band magnitude and $v_{circ}$ is the maximum rotational velocity of the disc in km\,s$^{-1}$. \citet{obre13} find

\ben
\begin{split}
M_K \sim~& \mathcal N[\mu=(-9.3\pm0.3) \log_{10} v_{circ}-(2.5\pm0.7),\\
 &\sigma=0.22\pm0.06],
\end{split}
\een

\noindent where all the fit parameters comfortably agree within quoted errors. The parameter errors in \citet{obre13} were recovered via bootstrap resampling of the data points, whereas \hf used the covariant matrix via the inverse of the parameter Hessian matrix. Using \hf we find marginally less constraint on the slope and intercept, but slightly more on the intrinsic scatter.

\subsection{3D Fundamental Plane}

\begin{figure}
	\centering
	\includegraphics[width=\columnwidth]{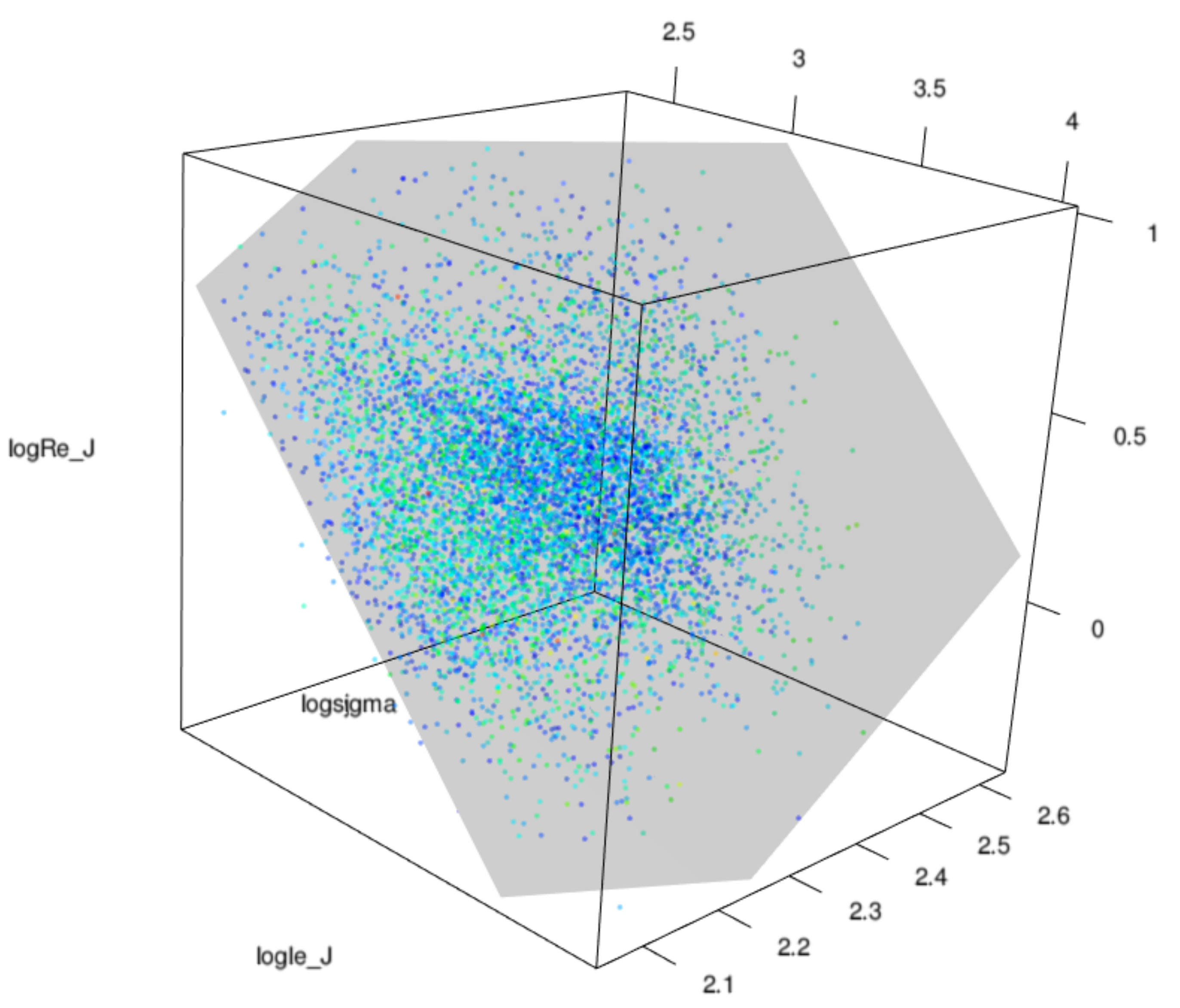}
	\includegraphics[width=\columnwidth]{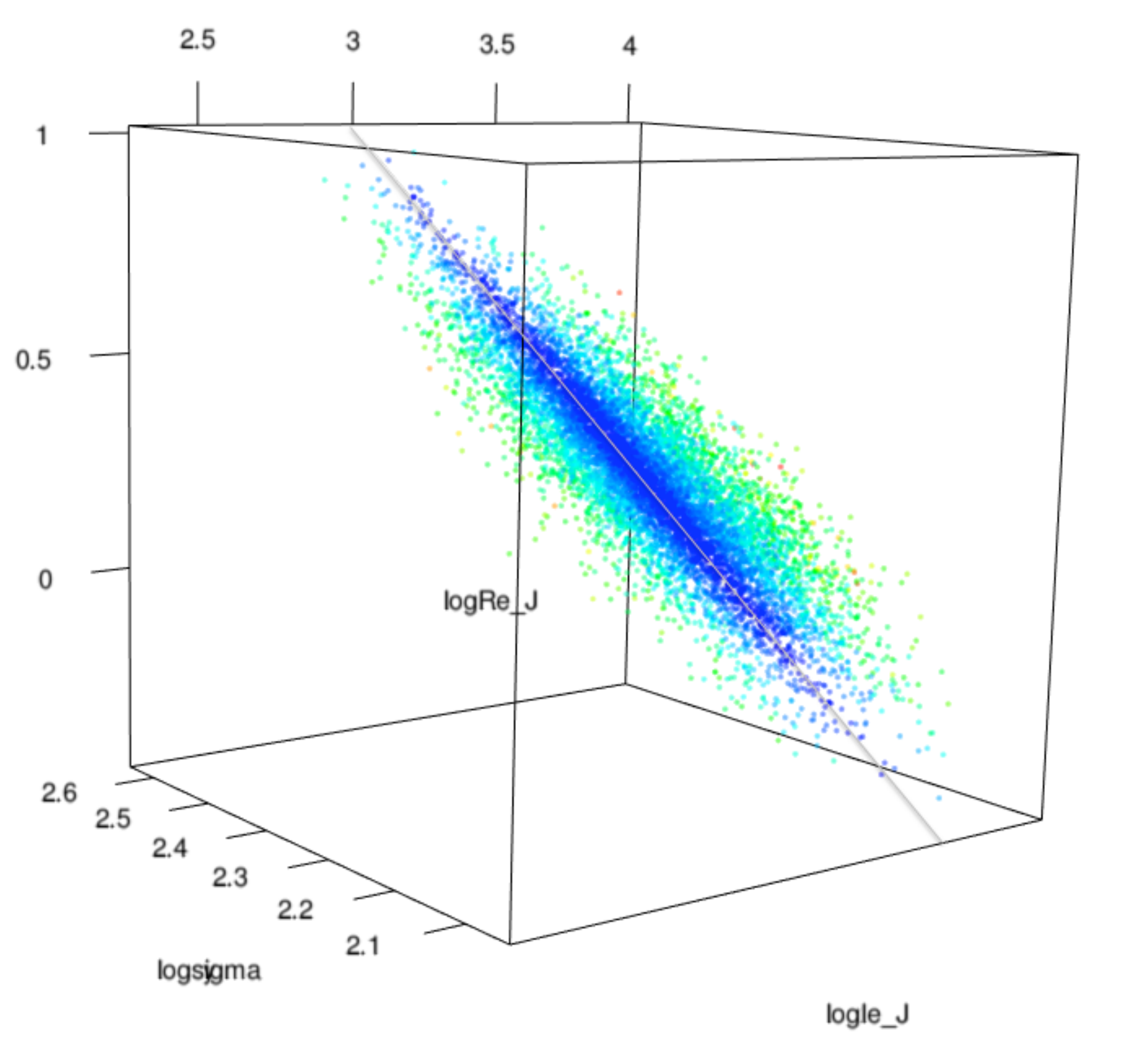}
	\caption{6dFGS Fundamental Plane data and \hf fit. The two panels are different orientations of the default plot output of the {\tt hyper.plot3d} function (accessed via the class specific {\tt plot} method) included as part of the \R \hf package, where the best-fit \hf generative model for the data is shown as a translucent grey 3D plane. The package function is interactive, allowing the user the rotate the data and the overlaid 3D plane to any desired orientation. See Figure \ref{fig_fitnoerror} for further details on this Figure.
	\label{fig_FP}}
\end{figure}

Moving to a 3D example, perhaps the most common application in the literature is the Fundamental Plane for elliptical galaxies \citep{fabe87,binn98}. This also offers a route to a redshift-independent distance estimator, but from a 3D relation rather than a 2D relation \citep[although an approximate 2D version also exists for elliptical galaxies via the Faber-Jackson relation;][]{fabe76}. In \hf we include 6dFGS data released in \citet{camp14}, where we fit the data and generate Figure \ref{fig_FP} with\\

\noindent {\tt
> data(FP6dFGS)\\
> plot(hyper.fit(FP6dFGS[,c("logIe\_J", "logsigma", "logRe\_J")], vars=FP6dFGS[,c("logIe\_J\_err", "logsigma\_err", "logRe\_J\_err")]\^{}2, weights=FP6dFGS[,"weights"])) \hfill \textrm{(Figure \ref{fig_FP})}
}\newline

This yields a Fundamental Plane relation of

\ben
\begin{split}
\log_{10} Re_J \sim~& \mathcal N[\mu= -(0.853 \pm 0.005) \log_{10} Ie_J \\& + (1.508 \pm 0.012) \log_{10}\sigma_{vel} - (0.42 \pm 0.03),\\&
 \sigma=0.060\pm0.001],
\end{split}
\een

\noindent where $Re_J$ (in kpc) is the effect radius in the $J$-band, $Ie_J$ (in mag~arcsec$^{-2}$) is the average surface brightness intensity within $Re_J$ and $\sigma_{vel}$ (in km\,s$^{-1}$) is the central velocity dispersion. \citet{mago12} find

\ben
\begin{split}
\log_{10} Re_J \sim~& \mathcal N[\mu= -(0.89 \pm 0.01) \log_{10} Ie_J \\ &+ (1.52 \pm 0.03) \log_{10}\sigma_{vel} - (0.19 \pm 0.01),\\&
 \sigma=0.12\pm0.01].
\end{split}
\een

It is notable that our fit requires substantially less intrinsic scatter than \citet{mago12}. It is difficult to assess the origin of this difference since the methods used to fit the plane differ between this work and \citet{mago12}, where the latter fit a 3D generative Gaussian with 8 degrees of parameter freedom, which is substantially more general than the 3D plane with only 4 degrees of parameter freedom that \hf used.

\subsection{3D Mass-Spin-Morphology Relation}

\begin{figure}
	\centering
	\includegraphics[width=\columnwidth]{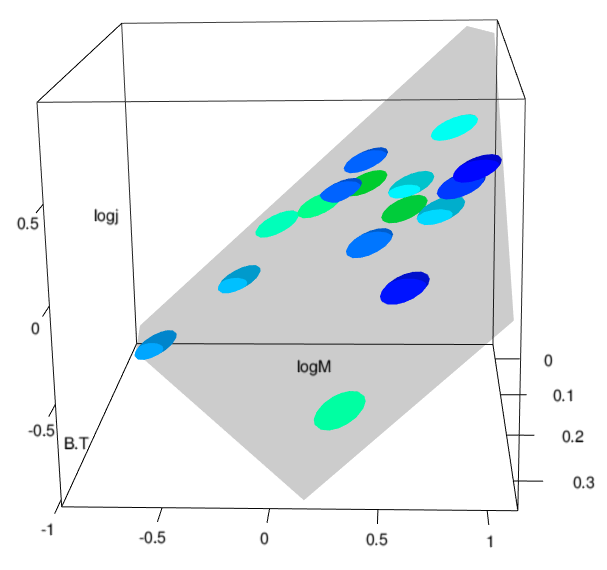}
	\includegraphics[width=\columnwidth]{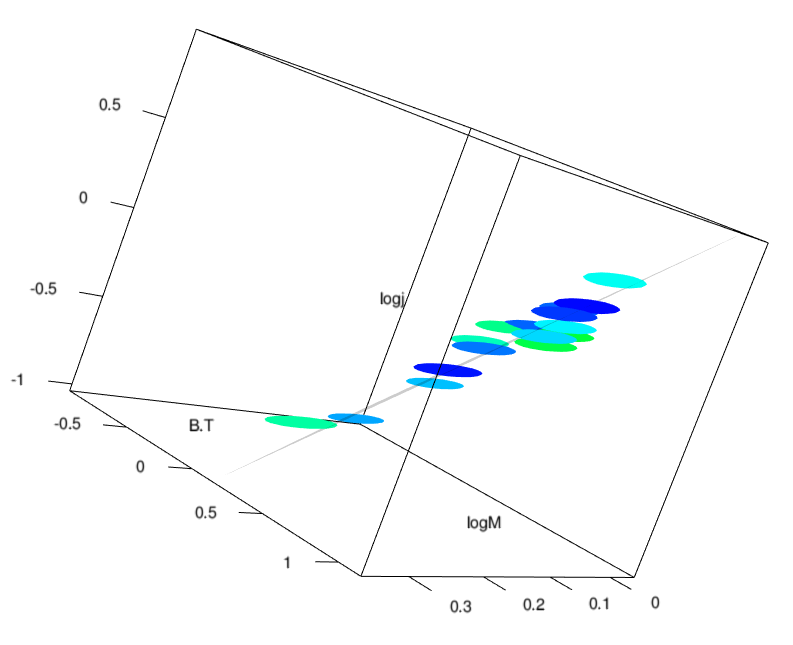}
	\caption{Mass-spin-morphology (MJB) data and \hf fit. The two panels are different orientations of the default plot output. All error ellipsoids overlap with the best-fit 3D plane found using \hf, implying that the generative model does not require any additional scatter. Indeed the observed data is unusually close to the plane, implying slightly overestimated errors. See Figure \ref{fig_fitnoerror} and \ref{fig_FP} for further details on this Figure.
	\label{fig_MJB}}
\end{figure}

The final example of \hf uses astronomical data from \citet{obre14}. They measured the fundamental relationship between mass (here the baryon mass (stars+cold gass) of the disc), spin (here the specific baryon angular momentum of the disc) and morphology (as measured by the bulge-to-total mass ratio). Hereafter this relation is referred to as the MJB relation. This dataset is interesting to analyse with \hf because it includes error correlation between mass and spin, caused by their common dependence on distance. In \hf we included the MJB data presented in \citet{obre14}. We can fit these data and generate Figure \ref{fig_MJB} with\\

\noindent {\tt
> data(MJB) \\
> plot(hyper.fit(X=MJB[,c("logM", "logj", "B.T")], covarray=makecovarray3d(MJB\$logM\_err, MJB\$logj\_err, MJB\$B.T\_err, MJB\$corMJ, 0, 0))) \hfill \textrm{(Figure \ref{fig_MJB})}
}\newline

%
%
%
%
%
%
%
%
%
%
%
%
%
%
%
%

This leads to a MJB relation of

\ben
\begin{split}
B/T \sim~& \mathcal N[\mu= (0.33 \pm 0.03) \log_{10} \mathcal M \\ &- (0.34 \pm 0.03) \log_{10} j - (0.04 \pm 0.01),\\
 &\sigma=0.002\pm0.006],
\end{split}
\een

\noindent where $B/T$ is the galaxy bulge-to-total ratio, $\mathcal M$ is the disc baryon mass in $10^{10}$ solar masses, and $j$ is the specific baryon angular momentum in $10^{-3}$\,kpc\,km\,s$^{-1}$ \citep[see][for details on measuring this quantity]{obre14}. \citet{obre14} find

\ben
\begin{split}
B/T \sim ~& \mathcal N[\mu= (0.34 \pm 0.03) \log_{10} \mathcal M \\
&- (0.35 \pm 0.04) \log_{10} j - (0.04 \pm 0.02),\\
&\sigma=0].
\end{split}
\een

The agreement between \hf and \citet{obre14} is excellent, with all parameters consistent within quoted 1$\sigma$ errors. Our results are consistent with zero intrinsic scatter within parameter errors. In fact, running the analysis using MCMC and the CHARM algorithm we used \hf to directly sample the likelihood of the intrinsic scatter being exactly 0. This is possible because we use the wall condition that any intrinsic scatter sample below 0 (which is non-physical) is assigned to be exactly 0. We find that 90.4\% of the posterior chain samples have intrinsic scatter of exactly 0 post application of the wall condition (which imposes a spike + slab posterior), which can be interpreted to mean that the data is consistent with $P(\sigma=0)=0.904$.

This agrees with the assessment in \citet{obre14}, where they conclude that the 3D plane generated does not require any additional scatter to explain the observed data, probably indicating that the errors have been overestimated\footnote{If the real generative model is a hyperplane with no scatter then we expect to observe $P(\sigma=0)=U(0,1)$, where $U$ is the uniform distribution. $P(\sigma=0)$ can be cast as the frequentist p-value, where the null hypothesis is a generative model with no intrinsic scatter. The posterior we form via the addition of the wall condition is formally not a Bayesian posterior. A fully rigorous approach would be to compare Bayes factors for a model with $\sigma=0$ and another model with $\sigma>0$. Using the wall condition posterior as described should alert the user to situations where such a model comparison is potentially necessary.}.

\section{Conclusion}

In this paper we have extended previous work \citep{kell07,hogg10,kell11} to describe the general form of the $D$ dimensional generative hyperplane model with intrinsic scatter. We have also provided the fully documented \hf package for the \R statistical programming language, available immediately for installation via github\footnote{github.com/asgr/hyper.fit}. A user friendly \Shiny web interface has also been developed and can be accessed at hyperfit.icrar.org. The \hf version released in conduction with this paper is v1.0.2, and regular updates and support are planned.

To make the package as user-friendly as possible we have provided examples for simple 2D datasets, as well as complex 3D datasets utilising heteroscedastic covariant errors taken from \citet{obre14}. We compared the published fits to the new \hf estimates, and in most cases find good agreement for the hyperplane parameters (not all original papers attempt to make an intrinsic scatter estimate). \hf offers users access to higher level statistics (e.g.\ the parameter covariance matrix) and includes simple class sensitive {\tt summary} and {\tt plot} functions to assist in analysing the quality of the generative model.

The ambition is for \hf to be regularly updated based on feedback and suggestions from the astronomy and statistics community. As such, in the long term the documentation included as part of the package might supersede some details provided in this paper.

\section{Future Work}

The current \R \hf package does not have a mechanism for handling censored data, and we currently implicitly assume a uniform population distribution along the hyperplane. Both of these issues can arise with astronomy data, where typically some fraction of sources might be left censored (i.e.\ below a certain value but uncertain where, and usually represented by an upper limit) and sampled from an unknown power-law type distribution (\hf allows for the description of known power-law selection functions, see Appendix \ref{subsection_selection}). A future extension for \hf will allow for multi-parameter censoring, i.e.\ where a data point might exist some subset of parameters (e.g. stellar mass and redshift) but only has an upper limit for others (e.g.\ star-formation rate and HI mass). \citet{kell07} allows for censoring of the response variable, but not multiple observables simultaneously. It is possible to account for such censoring using a hierarchical Bayesian inference software program, such as \BUGS, \JAGS or \Stan. An example \JAGS model of the \hf scheme is provided in Appendix \ref{sec_JAGS}.  We also aim to allow generic distribution functions along and perpendicular to the hyperplane including power-laws and Gaussian mixture models. Again, using a hierarchical Bayesian inference software program is one potential route to achieving this, but at the expense of generalisability and simplicity.

\section*{Acknowledgments}

Thank you to Joseph Dunne who developed the web tool interface for \hf which we now host at hyperfit.icrar.org. This was done as part of an extended summer student programme and required remarkably little direct supervision. Thank you to Ewan Cameron who assisted in describing the correct \JAGS 2D version of the \hf code presented in Appendix \ref{sec_JAGS}. Thank you finally to the members of the Astrostatistics FaceBook group\footnote{www.facebook.com/groups/astro.r} who contributed helpful comments at early stages of this work, and gave useful feedback on the \R \hf package prior to publication.

\bibliographystyle{mn2e}
\setlength{\bibhang}{2.0em}
\setlength\labelwidth{0.0em}


\appendix


\section{Intrinsic scatter estimator bias}
\label{subsection_bias}

A subtle point is that the {\it expectation of the most likely model is not generally identical to the true model}. Let us assume a fixed generative model described by Eq.~(\ref{eq_rhom}) with known parameters $\n^{\rm true}$ and $\sigma_\perp^{\rm true}$. From this model, we randomly draw a sample of $N$ points and estimate the most likely values, i.e.\ the {\it modes}, $\n^{\rm ML}$ and $\sigma_\perp^{\rm ML}$ by maximising Eq.~(\ref{eq_likelihood_coordfree}). This procedure is repeated ad infinitum, always with the same model and the same number of points, resulting in infinitely many different estimates $\n^{\rm ML}$ and $\sigma_\perp^{\rm ML}$. By definition, the {\it expectations} of these estimators are the ensemble-averages $\langle\n^{\rm ML}\rangle$ and $\langle\sigma_\perp^{\rm ML}\rangle$. The question is whether these expectations are identical to the input parameters $\n^{\rm true}$ and $\sigma_\perp^{\rm true}$. If so, the estimators are called unbiased, if not, they are biased. The mirror-symmetry of our generative model implies that $\n^{\rm ML}$ is an unbiased estimator\footnote{Note that $\n^{\rm ML}$ can become a biased estimator, when the data is drawn from the population with a non-uniform selection function, as discussed in Appendix~\ref{subsection_selection}.}, i.e.\ $\langle\n^{\rm ML}\rangle=\n^{\rm true}$.

\begin{figure}
	\centering
	\includegraphics[width=\columnwidth]{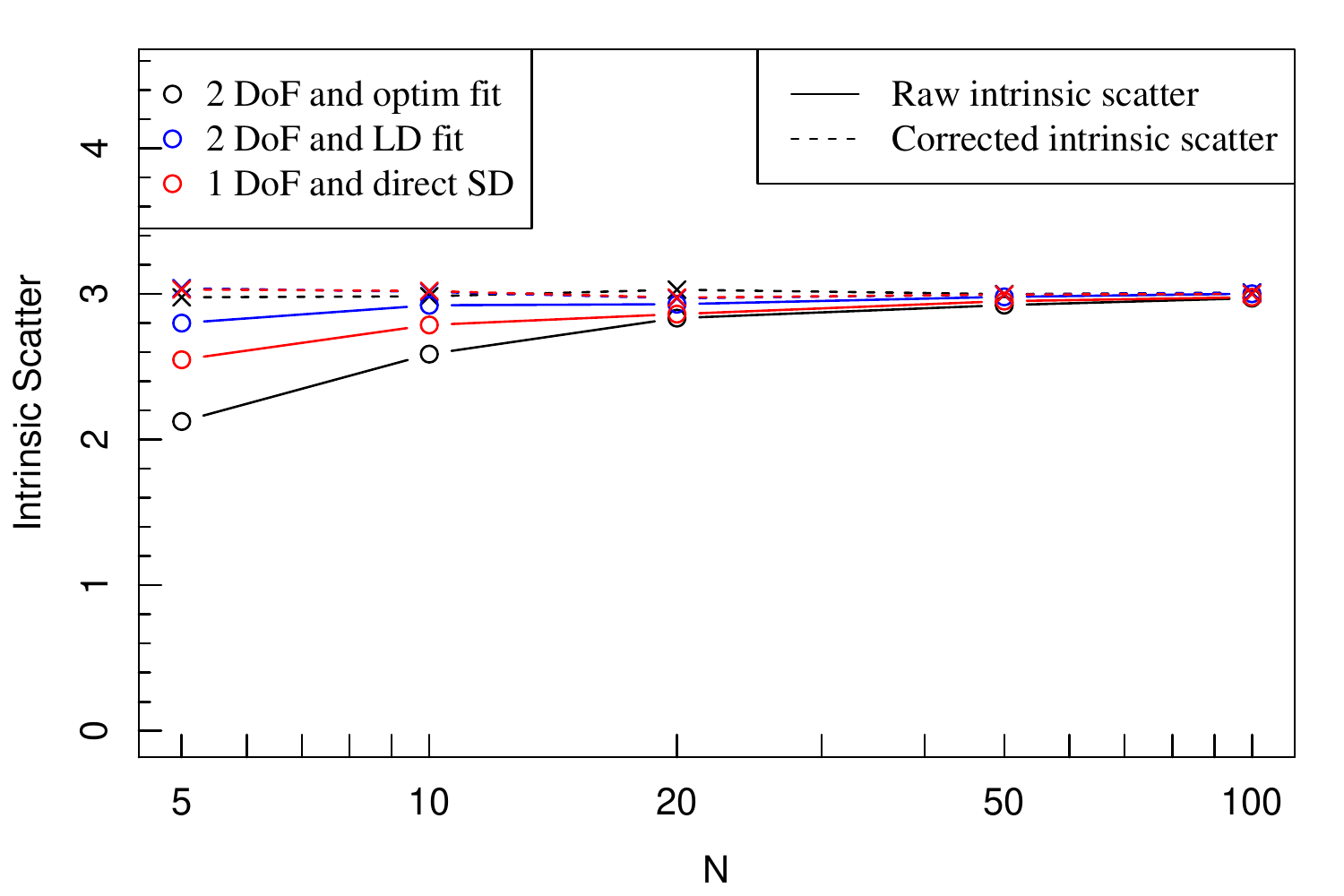}
	\caption{Convergence tests for simulated generative hyperplane datasets. The solid lines show the raw mean intrinsic scatter for different types of generative model and fitting, and the dashed lines show the intrinsic scatter once the appropriate combination of bias and sample--population corrections have been applied. In all cases the generative model was simulated with an intrinsic scatter for the population equal to 3.
	\label{fig_convtest}}
\end{figure}

By contrast, the intrinsic scatter is generally biased, i.e.\ $\langle\sigma_\perp^{\rm ML}\rangle\neq\sigma_\perp^{\rm true}$. There are two reasons for this bias. Firstly, maximising the likelihood function $\L$ finds the most likely generative model of a {\it sample} of $N$ data points, which is {\it not} generally the most likely model of the {\it population}, i.e.\ the imaginary infinite set, from which the $N$ points were drawn at random. For example, consider a two-dimensional population described by a straight line with some intrinsic scatter. From this population we draw a sample of two points. No matter which two points we choose, they can always be fitted by a straight line without scatter. Hence the most likely sample model, maximising Eq.~(\ref{eq_likelihood_coordfree}), will have zero intrinsic scatter, even if the population model has non-zero scatter. The transition from the most likely model of the sample to the most likely model of the population is achieved by the so-called Bessel correction,
\be
	\tilde{\sigma}_\perp^{\,2} = \frac{N}{N-D}~\sigma_\perp^{\,2},
	\label{eq_bias1}
\ee
where $D$ is the number of parameters of the hyperplane (also referred to as the `degrees of parameter freedom' or simply `degrees of freedom'), which equals the number of dimensions. In the special case of Gaussian data with no errors ($\C_i\equiv0$), the variance estimator $\tilde{\sigma}_\perp^{\,2}$ is unbiased, i.e.\ $\langle(\tilde{\sigma}_\perp^{\rm ML})^2\rangle=(\sigma_\perp^{\rm true})^2$. In the general case, one can estimate the population variance by computing the mean of its marginal posterior PDF. This is often achieved using a Markov Chain Monte Carlo (MCMC) approach, but note that many MCMC algorithms introduce new biases in the way they determine when a Markov Chain becomes stable \citep{cowl99}.

Secondly, the non-linear relation between $\tilde{\sigma}_\perp^{\,2}$ and $\tilde{\sigma}_\perp$ skews the posterior PDF in such a way that the expectation of the mode $\tilde{\sigma}_\perp^{\rm ML}$ is generally smaller than $\sigma_\perp^{\rm true}$. In the case of data with no errors ($\C_i\equiv0$), an unbiased estimator can be derived from Cochran's theorem,
\be\label{eq_bias2}
	\tilde{\tilde{\sigma}}_\perp \!=\! \sqrt{\frac{N\!-\!D}{2}}\frac{\Gamma\big(\frac{N-D}{2}\big)}{\Gamma\big(\frac{N-D+1}{2}\big)}\tilde{\sigma}_\perp \!=\! \sqrt{\frac{N}{2}}\frac{\Gamma\big(\frac{N-D}{2}\big)}{\Gamma\big(\frac{N-D+1}{2}\big)}\sigma_\perp.
\ee
This estimator then satisfies $\langle\tilde{\tilde{\sigma}}_\perp\rangle=\sigma_\perp^{\rm true}$. \hf includes the Bessel and Cochran corrected versions of the intrinsic scatter (where appropriate). Figure \ref{fig_convtest} gives three example fits for different DoF and different samples $N$. The black and the red solid lines are direct maximum likelihood estimation fits. These are the most biased initially, requiring both a Bessel and Cochran correction to give the dashed unbiased lines. The blue line shows a MCMC expectation estimation, and as such only requires the Cochran correction to give the unbiased dashed line. These corrections do not properly account for the presence of heteroscedastic errors, but they should offer the {\it minimum} appropriate correction to apply. Bootstrap resampling offers a window to properly correct for such data biases, but this can be extremely expensive to compute.

\section{Extension to non-uniform selection functions}
\label{subsection_selection}

We now consider the more general situation, where the population is not sampled uniformly, but with a probability proportional to $\rho_s(\x)$, known as the selection function in astronomy. In this case, the effective PDF $\rho_{m'}(\x)$ of the generative model \emph{and} the selection function is the product $\rho_m(\x)\rho_s(\x)$, renormalised along the direction perpendicular to $\H$, $\int_{-\infty}^{\infty}dl\,\rho_{m'}(l\hat{\n})=1$. The renormalisation is crucial to avoid a dependence of the normalisation of $\rho_{m'}(\x)$ on the fitted model parameters. The effective PDF reads
\be
 	\rho_{m'}(\x) = \frac{\rho_m(\x)\rho_s(\x)}{\int_{-\infty}^{\infty}dl\,\rho_m(l\hat{\n})\rho_s(l\hat{\n})}
	\label{eq_effectivePDF}
\ee
The likelihood of point $i$ then becomes 
\be
	\L_i = \int_\RD\!\!d\x\,\rho(\x_i|\x)\rho_{m'}(\x).
	\label{eq_Li_selection}
\ee

Analytical closed-form solutions of Eq.~(\ref{eq_Li_selection}) can be found for some functions $\rho_s(\x)$. To illustrate this, let us consider the case of an exponential selection function
\be
	\rho_s(\x) = e^{\k\t\x},
	\label{eq_rhos}
\ee
where $\k$ is the known (and fixed) sampling gradient, i.e.\ in the direction of $\k$ the probability of drawing a point from the population increases exponentially. This selection function is important in many fields of science, since it represents a power-law selection function in the coordinates $\tilde{\x}$, if $\x$ is defined as $\x=\ln\tilde{\x}$. This can be seen from
\be
	e^{\k\t\x} = \prod_{j=1}^D e^{k_j x_j} = \prod_{j=1}^D\big(e^{x_j}\big)^{k_j} = \prod_{j=1}^D\tilde{x}_j^{k_j},
\ee
which is a generic $D$-dimensional power-law.

Substituting $\rho_m(\x)$ and $\rho_s(\x)$ in Eq.~(\ref{eq_effectivePDF}) for Eqs.~(\ref{eq_rhom}) and (\ref{eq_rhos}) solves to
\be
	\rho_{m'}(\x) = \frac{1}{\sqrt{2\pi\sigma_\perp^2}}\,e^{-\frac{(\hat{\n}\t[\x-\sigma_\perp^2\k]-n)^2}{2\sigma_\perp^2}},
	\label{eq_rhomp}
\ee
In analogy to Eq.~(\ref{eq_simpleLi}), the likelihood terms then become
\be
	\L_i = \int_\RD\!\!d\x\,\rho(\x_i|\x)\rho_{m'}(\x) = \frac{1}{\sqrt{2\pi s_{\perp,i}^2}}\,e^{-\frac{(\hat{\n}\t\xx_i-n)^2}{2s_{\perp,i}^2}},
	\label{eq_powerlawLi}
\ee
where $s_{\perp,i}^2=\sigma_\perp^2+\hat{\n}\t\C_i\hat{\n}$ and $\xx_i=\x_i-\sigma_\perp^2\k$. Up to an additive constant, the total likelihood function $\ln\L=\sum_{i=1}^{N}\ln\L_i$ reads
\be
	\ln\L = -\frac{1}{2}\sum_{i=1}^{N} \left[\ln(s_{\perp,i}^2)+\frac{(\hat{\n}\t\xx_i-n)^2}{s_{\perp,i}^2}\right].
	\label{eq_L_selection}
\ee
This likelihood is identical to that of the uniform case (Eq.~(\ref{eq_likelihood_coordfree})), as long as we substitute the positions $\x_i$ for $\xx_i$. Upon performing this substitution, all results of Sections \ref{subsection_generaluniform} to \ref{subsection_twodim} remain valid in the case of a power-law sampling. Incidentally, it is easy to see that $\xx_i=\x_i$ if the data are sampled uniformly ($\k=0$).

The \hf package allows for the addition of such a power-law selection function via the $\tt{k.vec}$ vector input argument in the {\tt hyper.like} and {\tt hyper.fit} functions.

\section{Hierarchical implementation of 2D Hyper-Fit using JAGS}
\label{sec_JAGS}

Over recent years a number of hierarchical Bayesian inference software programs have become popular, e.g. Bayesian inference Using Gibbs Sampling (\BUGS\footnote{www.mrc-bsu.cam.ac.uk/software/bugs/}), Just Another Gibbs Sampler (\JAGS\footnote{mcmc-jags.sourceforge.net/}) and \Stan\footnote{mc-stan.org}. They are all somewhat similar in terms of the modelling language, with an emphasis on constructing complex hierarchical models with relative ease \citep{gelm06}. An example 2D implementation of the generative \hf model can be constructed in \JAGS using the following model specification:

\begin{lstlisting}
model {
  for (i in 1:N) {
    xyobs[1:2,i] ~ dmnorm(xytrue[1:2,i],
                   errors[1:2,1:2,i])
  }
  for (i in 1:N) {
    errors[1,1,i] <- 1/xsd[i]^2/(1-rho[i]^2)
    errors[2,2,i] <- 1/ysd[i]^2/(1-rho[i]^2)
    errors[1,2,i] <- rho[i]/xsd[i]/
                    ysd[i]/(1-rho[i]^2)
    errors[2,1,i] <- rho[i]/xsd[i]/
                    ysd[i]/(1-rho[i]^2)
  }
  for (i in 1:N) {
    xytrue[1,i] <- distfromorigin[i]*
                  cos(angle*0.01745)-
                  sin(angle*0.01745)*scat[i]
    xytrue[2,i] <- distfromorigin[i]*
                  sin(angle*0.01745)+
                  cos(angle*0.01745)*scat[i]+
                  originy
  }
  for (i in 1:N) {
    distfromorigin[i] ~ dunif(-1000,1000)
    scat[i] ~ dnorm(0,1/intrinvar)
  }
  angle ~ dunif(-90,90)
  originy ~ dunif(-1000,1000)
  intrinvar ~ dunif(0,10000)
}
\end{lstlisting}

In the above 2D model the observed $x$-$y$ data is contained in the $2 \times N$ matrix $\tt{xytrue}$, the $x$ errors are in the $N$ vector $\tt{xsd}$, the $y$ errors are in the $N$ vector $\tt{ysd}$, and the error correlations between $x$ and $y$ are in the $N$ vector $\tt{rho}$. The rotation angle of the generative model, the offset of the model from the origin, and the intrinsic variance of the model are all sampled from uniform prior distributions in this example. The model specification using \Stan or other \BUGS style software programs is very similar in construct to the above.

Compared to the \hf software, the \JAGS modelling solution converges slowly (factors of a few slower for equally well sampled posteriors) and tends to suffer from poor mixing between highly correlated parameters. This is a common problem with Gibbs sampling, and in practice the typical user can gain confidence in their fit solution using \hf by attempting a number of the available routines, where e.g.\ Slice Sampling is an empirical approximation to true Gibbs sampling. Fully hierarchical based solutions (e.g.\ \BUGS, \JAGS and \Stan) are also hard to extend into arbitrary dimensions, and in general need to be carefully hand-modified for bespoke datasets. However, these programs offer clear advantages in elegantly handling left or right censored data, with very little modification to the above specified model required.

\end{document}